\documentclass[final]{jfm-navid}
\usepackage{newtxtext}
\usepackage{newtxmath}
\usepackage{amsbsy}
\usepackage{epsfig}
\usepackage{graphicx}

\usepackage{amssymb, amsmath}
\usepackage[mathscr]{eucal}
\usepackage{natbib}
\usepackage{comment}
\usepackage{microtype}
\usepackage[usenames, dvipsnames]{xcolor}

\usepackage{wry_abbreviationsHC}

\usepackage{showlabels}

\usepackage{hyperref}
\hypersetup{
  breaklinks = true,
  colorlinks = true,
  linkcolor  = blue,
  citecolor  = Purple,
  urlcolor   = Purple,
  pdfauthor  = {Navid C. Constantinou, Cesar B. Rocha, Stefan G. Llewellyn Smith, and William R. Young},
  pdftitle   = {Nusselt number scaling in horizontal convection},
 }

\newcommand{\ssquare}{\scalebox{0.6}{$\square$}}
\newcommand{\sbsquare}{\scalebox{0.6}{$\blacksquare$}}

\newcommand{\ellx}{\ell_x}
\newcommand{\elly}{\ell_y}

\newcommand{\h}{h}

\newcommand{\NS}{\mathit{NS}}
\newcommand{\FS}{\mathit{FS}}

\newcommand{\chidiff}{\chi_{\mathrm{diff}}}
\newcommand{\bdiff}{b_{\mathrm{diff}}}

\newcommand{\bs}{\ensuremath{b_{\mathrm{s}}}}

\newcommand{\bstar}{\ensuremath{b_{\star}}}

\newcommand{\Nu}{\mbox{\textit{Nu}}}

\newcommand{\Nus}{\Nu_{\mathrm s}}

\newcommand{\Ra}{\mbox{\textit{Ra}}}
\renewcommand{\Re}{\mbox{\textit{Re}}}

\renewcommand{\Pr}{\mbox{\textit{Pr}}}

\renewcommand{\sech}{\mathop{\rm sech}\nolimits}
\renewcommand{\cosech}{\mathop{\rm cosech}\nolimits }

\newcommand{\dab}{\delta_b}
\newcommand{\dau}{\delta_u}

\newcommand{\dafour}{\delta_{\scriptscriptstyle{1/4}}}
\newcommand{\dafive}{\delta_{\scriptscriptstyle{1/5}}}

\newcommand{\Kfour}{K_{\scriptscriptstyle{1/4}}}
\newcommand{\Kfive}{K_{\scriptscriptstyle{1/5}}}

\newcommand{\etaK}{\eta_{\mathrm{K}}}

\newcommand{\etaB}{\eta_{\mathrm{B}}}

\renewcommand{\NS}{\textcolor{Red}{NS}}
\renewcommand{\FS}{\textcolor{blue}{FS}}
\newcommand{\twoDNS}{\textcolor{Red}{2DNS}}
\newcommand{\threeDNS}{\textcolor{Red}{3DNS}}
\newcommand{\twoDFS}{\textcolor{blue}{2DFS}}
\newcommand{\threeDFS}{\textcolor{blue}{3DFS}}

\usepackage{printlen}

\begin{document}

\title[Horizontal convection $\Nu$ scaling]{Nusselt number scaling in horizontal convection}

\author{Navid~C.~Constantinou\aff{1}
\corresp{\email{navid.constantinou@anu.edu.au}},
Cesar~B.~Rocha\aff{2},
Stefan~G.~Llewellyn~Smith\aff{3, 4},
\and William~R.~Young\aff{4}}

\affiliation{
  \aff{1}Research School of Earth Sciences \& ARC Centre of Excellence for Climate Extremes, Australian National University, Canberra, ACT 2601, Australia
  \aff{2}Instituto Oceanográfico, Universidade de São Paulo, São Paulo, SP 05508-120, Brasil
  \aff{3}Department of Mechanical and Aerospace Engineering, University of California San Diego, La Jolla, CA 92093-0411, USA
  \aff{4}Scripps Institution of Oceanography, University of California San Diego, La Jolla, CA 92093-0213, USA
}

\shortauthor{N.~C.~Constantinou, C.~B.~Rocha, S.~G.~L.~Smith, and W.~R.~Young}

\pubyear{}
\volume{}
\pagerange{}

\maketitle

\begin{abstract}

We report a numerical study of horizontal convection (HC) at Prandtl number $\Pr = 1$, with both  no-slip and free-slip boundary conditions. We obtain 2D and 3D solutions and determine the relation between the Rayleigh number $\Ra$ and the Nusselt number $\Nu$. In 2D we vary $\Ra$ between $0$ and $10^{14}$. In the range $10^6 \lessapprox \Ra \lessapprox 10^{10}$ the $\Nu$--$\Ra$ relation is $\Nu \sim \Ra^{1/5}$. With $\Ra$ greater than about $10^{11}$ we find a 2D regime with $\Nu \sim Ra^{1/4}$ over three decades, up to the highest 2D~$\Ra$. In 3D, with maximum $\Ra = 10^{11.5}$, we find only $\Nu \sim \Ra^{1/5}$. These results apply to both free slip and no slip boundary conditions. The $\Nu \sim Ra^{1/4}$ regime has a double boundary layer (BL): there is a thin BL with thickness $\sim \Ra^{-1/4}$ nested inside a thicker BL with thickness $\sim \Ra^{-1/5}$. The $\Ra^{-1/4}$ BL thickness, which determines $\Nu$, coincides with the Kolmogorov and Batchelor scales of HC.

Numerical and theoretical results indicate that 3D HC is qualitatively and quantitatively similar to 2D HC. At the same $\Ra$, the 3D $\Nu$ exceeds the 2D $\Nu$ by less than  $20$\%, i.e., there is very little 3D enhancement of heat transport. Boundary conditions are more important than dimensionality: the  2D free-slip solutions have larger $\Nu$ than 3D no-slip solutions. Using the  mechanical energy power integral of HC we show that the mean square vorticity of 3D HC is nearly equal to that of 2D HC at the same $\Ra$. Thus vorticity amplification by strain-mediated vortex stretching does not operate in 3D HC.
\end{abstract}

\section{Introduction}

Horizontal convection (HC) is convection driven by imposing non-uniform heating and cooling along a single horizontal surface, such as the top of a rectangular enclosure.
There is no flux of heat through the other boundaries \citep{HG08}.
Oceanography is an important motivation for consideration of HC \citep{S08, R65, CGH06}, and in that connection the dependence of horizontal heat transport on the strength of buoyancy forcing applied at the ocean surface is a prime question.
Buoyancy forcing is quantified via the horizontal-convective Rayleigh number, $\Ra$, and horizontal heat transport by a suitably defined Nusselt number $\Nu$ \citep{RocNussDef}.


HC is an interesting counterpoint to Rayleigh-B\'enard convection because HC buoyancy transport in the interior of the domain cannot be easily interpreted as the vertical motion of thermal plumes.
Instead, in HC heat enters the fluid where the non-uniform heated surface is hotter than average and exits where it is colder.
This horizontal transport is associated with a prominent boundary layer (BL) adjacent to the non-uniform surface.

\begin{figure}
  \centering
  \includegraphics[width=0.9\textwidth]{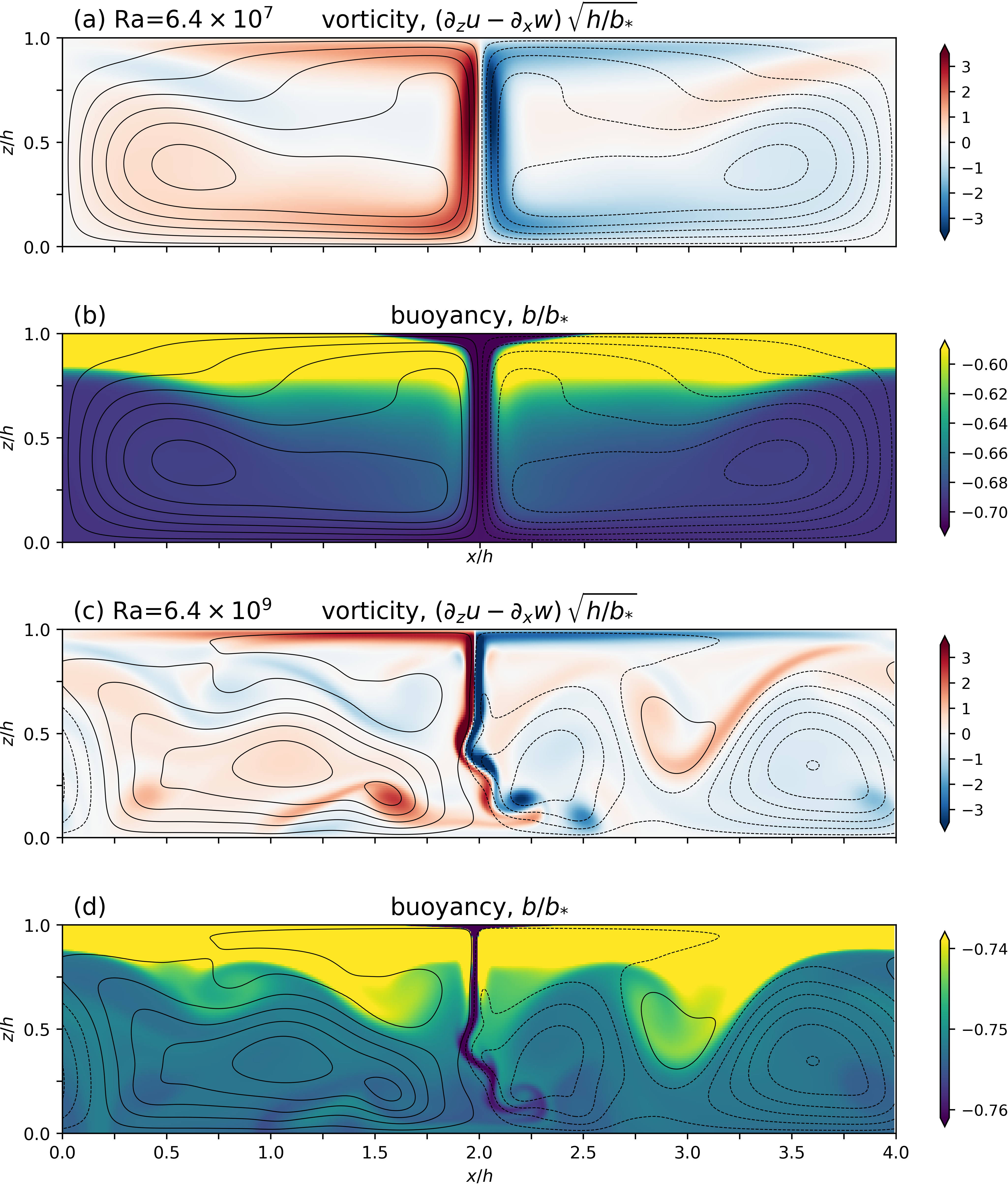}
  \caption{Snapshots of 2D free-slip (\twoDFS) HC at $\Ra = 6.4 \times 10^{7}$ in panels (a) and~(b) and $\Ra = 6.4 \times 10^{9}$ in (c) and~(d); $\Pr = 1$. Contours are streamlines. At the top surface $-1 \leq b /\bstar \leq +1$. The narrow range of the buoyancy color scale in (b) and~(d) makes the very small interior buoyancy variations visible. The sinusoidal surface buoyancy in~\eqref{sbuoy8} defines the buoyancy scale $\bstar$; $h$ is the layer depth.}
  \label{Fig1}
\end{figure}

\subsection{Rossby scaling}

\begin{table}
\begin{center}
\begin{tabular}{lcccccc}
Citation & $\Pr$  & $\ell_x / h$ & $\text{max} \, \Ra$&  BC &forcing  & $p$\\
\hline
\\
\cite{R98} &  $10$ & 1 & $10^8$ & \FS &  linear  &$\frac{1}{5}$\\
\\
\cite{SKB04}   & $\lbrace \frac{1}{5}, \cdots,  4 \rbrace$ & 1 &  $\approx 10^{9}$ & \FS & sin & $\frac{1}{5}$ \\
\\
\cite{CWHL08} & $\infty$ & 1 & $10^{10}$ & \FS, \NS & linear & $\frac{1}{5}$ \\
  \\
\cite{SK2011} & 6.14 & $\lbrace1, \, 1.6,\, 6.25 \rbrace  $ & $10^{12}$ & \NS & linear & $\frac{1}{5}, \, > \frac{1}{5}?$\\
\\
\cite{RH2019}  & $\infty$ & 1 & $10^{10}$ & \FS & linear & $\frac{1}{5}$\\
  & $\infty$ & 1 & $10^{10}$ & \FS & various & $\frac{1}{5}, \, \frac{1}{4}$?\\
\\
\cite{tsaiSheard20}  & 6.14 & 6.25 & $10^{12.5}$ & \NS &  linear & $\frac{1}{5}, \, \frac{1}{4}$ \\
  & 6.14 & 6.25 & $10^{12.5}$ & \NS & various & $\frac{1}{5}$\\
  \\
This work & $1$ & 4 & $10^{13.8}$ & \FS, \NS & sin & $\frac{1}{5} \com \frac{1}{4}$\\
\end{tabular}
\end{center}
\caption{Summary of seven studies of $\Nu$--$\Ra$ HC scaling using 2D direct numerical simulations. Prandtl number $\Pr$ is fixed at the values in column~2 and $Ra$ is varied. In the final column $p$ is the exponent in~\eqref{alpBet333}. The ``forcing'' column refers to the surface buoyancy profile $\bs(x)$: ``linear'' is $\bs(x) \propto x$;  ``sin'' is sinusoidal; ``various'' is other $\bs(x)$ profiles. In the boundary condition (BC) column \NS{} is no slip and \FS{} is free slip. \cite{SKB04} used seven $Pr$'s between $0.2$ and $4$. \cite{SK2011} claim an elevation in $p$ above $1/5$ for $\Ra \gtrsim 10^{10}$. \cite{RH2019} indicate that depending on the surface forcing profile $p = 1/4$ might be a better fit than $p = 1/5$ in the range $10^8 \lesssim \Ra \leq 10^{10}$.}
\label{Table1}
\end{table}

The oldest result for the high-$\Ra$ variation of horizontal-convective $\Nu$ is the scaling law of \cite{R65},
\beq
  \Nu \sim \Ra^{1/5}\,  \Pr^0 \com
  \label{alpBet3}
\eeq
where $\Pr$ is the Prandtl number.
(Parameters $\Ra$, $\Pr$, and $\Nu$ are defined in section~\ref{form}.)
Rossby was motivated by experiments done using five values of $\Pr$ between $13$ and $8\, 500$.
Rossby's reasoning leading to~\eqref{alpBet3} (reviewed in section~\ref{RossbySec}) assumes  visco-diffusive BL dynamics, justified by the assumption that $\Pr \gg 1$.

``Rossby scaling'' often means the exponent $p=1/5$ in~\eqref{alpBet3} without reference to $\Pr^0$ and without mentioning Rossby's requirement that $\Pr \gg 1$.
Rossby was aware of $\Pr^0$:  in \cite{R98} he demonstrated the weak dependence of $\Nu$ on $\Pr$ in the range $3 \leq \Pr \leq 100$ (with fixed  $\Ra$'s between $10^4$ and  $10^8$).
In view of alternative scaling laws discussed below, all of which have the form, 
\beq
  \Nu \sim \Ra^{p}\,  \Pr^q \com
  \label{alpBet333}
\eeq
we reserve the term ``Rossby scaling'' for $(p, q)=(1/5, 0)$ with $\Pr \gg 1$.

\subsection{Non-Rossby scaling with  $p=1/5$}
 
At moderately large $\Ra$, the exponent $p=1/5$ is supported by both laboratory work and numerical solutions \citep{R98, M04, SKB04, WH2005, CWHL08, SK2011, IV12}.
In tables~\ref{Table1} and~\ref{Table2} we summarize various $\Nu$--$\Ra$ scaling laws claimed on the basis of 2D and 3D numerical simulations of HC: $p=1/5$ is the most common exponent of $\Ra$.

The exponent  $p=1/5$ occurs so frequently, and in such different parameter settings,  that  the necessity of Rossby's $\Pr \gg 1$ scaling argument is questionable.
For example, figure~\ref{Fig1} shows two $\Pr = 1$ numerical solutions both of which are   in the same  $p=1/5$ scaling regime (see section~\ref{RaNu}).
The $\Pr = 0.1$ points in figure~2(a) of \cite{ReiterShish2020} provide an example of $p=1/5$ scaling in a case with $\Pr \ll 1$.

 \cite{G14} make the important point that the exponent $1/5$ also arises with a non-Rossby balance between inertia and buoyancy in the surface BL.
 This balance leads to the  non-Rossby scaling
 \beq
 \Nu \sim \Ra^{1/5} \Pr^{1/5}\per
 \label{alt2}
 \eeq
\cite{Shish16} propose a $\Pr \ll 1$  non-Rossby scaling argument resulting in
 \beq
  \Nu \sim \Ra^{1/5} \Pr^{1/10}\per
  \label{alt3}
 \eeq
The scaling arguments leading to~\eqref{alpBet3}, \eqref{alt2} and~\eqref{alt3} are reviewed in  section~\ref{RossbySec} and put into a unified framework.
The results in~\eqref{alt2} and~\eqref{alt3} may help explain the tenacity of $1/5$ as the primary exponent $p$ in cases removed from Rossby's $\Pr \gg 1$ scenario.

\begin{table}
\begin{center}
\begin{tabular}{lccccccc}
Citation & $\Pr$  & $\ell_x / h$ & $\ell_y / h$ & $\text{max} \, \Ra$&  BC &forcing  & $p$\\
\hline
\cite{G14}& 5 &$6\frac{1}{4}$ & $\frac{1}{4} $ &  $10^{11.8}$ & \NS & pwc & $\frac{1}{5}$  \\ 
\\
\cite{ReiterShish2020}   & $ \lbrace 1 , \,  10 \rbrace$ & $10$ & $1$  & $10^{12.5} $ & \NS & pwc & $\frac{1}{4} \com  \frac{1}{5} \com\approx \frac{1}{4}$    \\
  & $\frac{1}{10}$ &  $10$ &  $1$  & $\sim 10^{11.7} $ & \NS & pwc & $\frac{1}{5}$ \\
\\
This work & $1$ & $4$ & $1$  & $10^{11.5}$ & \FS, \NS & sin  & $\frac{1}{5} $ 
\end{tabular}
\end{center}
\caption{Summary of three  studies of $\Nu$--$\Ra$ HC scaling using 3D direct numerical simulations. Prandtl number $\Pr$ is fixed at the values in column~2 and $Ra$ is varied. In the final column $p$ is exponent in~\eqref{alpBet333}. The ``forcing'' column refers to the surface buoyancy profile $\bs(x)$: ``pwc'' denotes a piecewise constant surface buoyancy  profile and ``sin'' is the  sinusoidal profile in~\eqref{sbuoy8}.}
\label{Table2}
\end{table}

 
\begin{figure}
  \centering
  \includegraphics[width = 0.9 \textwidth]{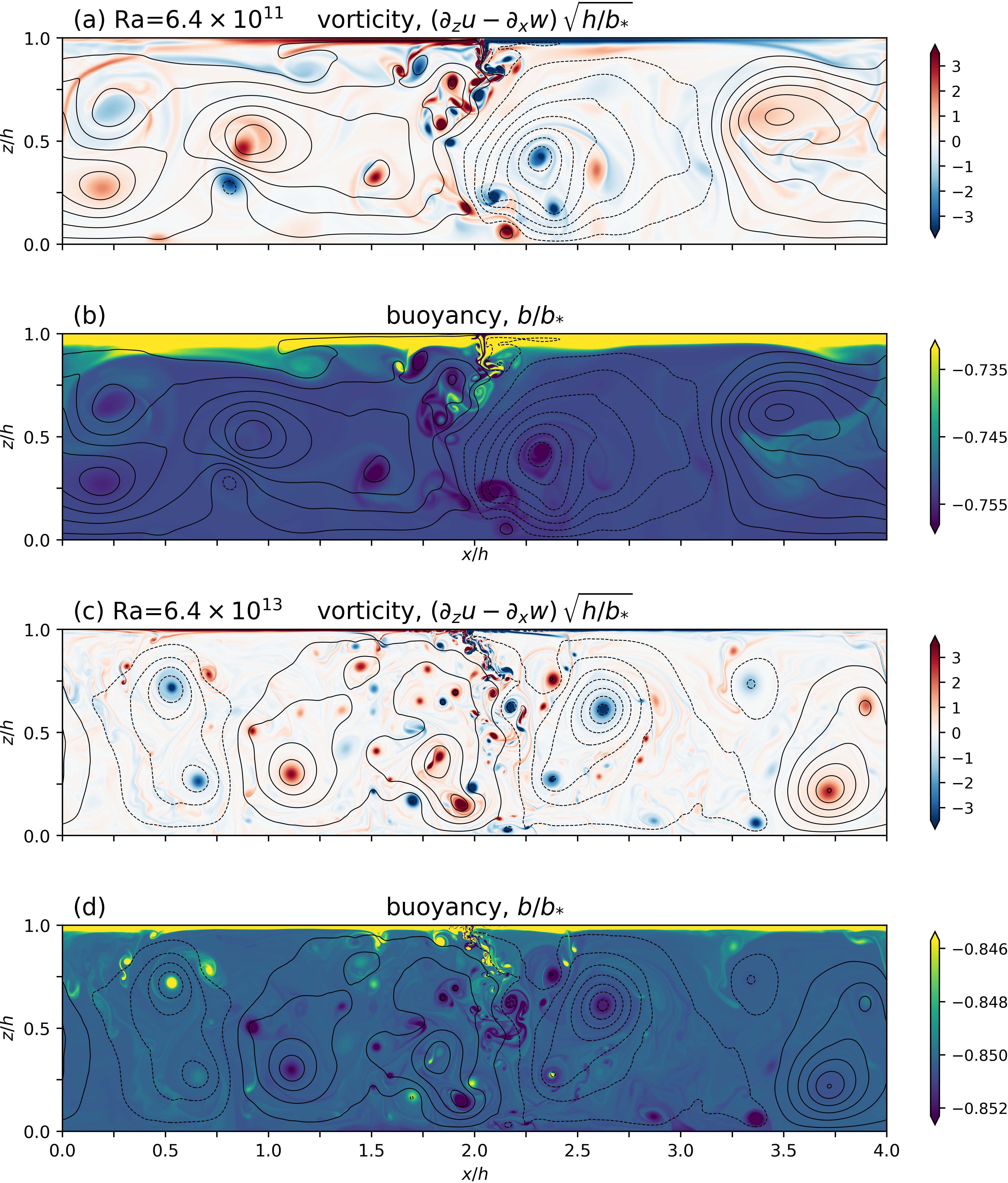}
  \caption{Snapshots of \twoDFS{} solutions at $\Ra = 6.4 \times 10^{11}$ in panels~(a) and~(b) and $\Ra = 6.4 \times 10^{13}$ in (c) and~(d); $\Pr = 1$. Contours are streamlines. This illustration uses true aspect ratio so that axisymmetric vortices look circular. The buoyancy color scale is narrow in order to reveal the small interior buoyancy variations, which are localized within the cores of axisymmetric vortices. Across the six-decade range of $\Ra$ in figures~\ref{Fig1} and~\ref{Fig2} vorticity scales with $\sqrt{\bstar/h}$.}
  \label{Fig2}
\end{figure}

\subsection{Higher $\Ra$ and $p=1/4$}

At moderately large $\Ra$,  $10^8 \lesssim \Ra \lesssim10^{10}$, \citet{tsaiSheard20} find $p=1/5$.
But in the range $10^{10} \lesssim \Ra \lesssim 10^{14}$, and provided that the imposed surface buoyancy varies linearly with the horizontal coordinate $x$, \citet{tsaiSheard20} report the scaling $p=1/4$.
In other words, as $\Ra$ increases at fixed $\Pr$  the primary exponent $p$ in~\eqref{alpBet333} increases from $1/5$ to $1/4$.
In section~\ref{RaNu} we report the same result: increasing $\Ra$, with $\Pr=1$, results in an increase in $p$ from $1/5$ to~$1/4$.
The solutions in figure~\ref{Fig1} are in the $p=1/5$ scaling regime while those in figure~\ref{Fig2} are in the  regime with $p=1/4$.
This increase in $p$ with  $\Ra$ is accompanied by a qualitative changes in the structure of HC.
Comparing figure~\ref{Fig2} with figure~\ref{Fig1} there is a striking transition from laminar flow  in figure~\ref{Fig1} to 2D turbulence in figure~\ref{Fig2}.
The axisymmetric vortices  in figure~\ref{Fig2} indicate that interior dynamics is close to  that of a vortex gas characteristic of freely evolving unstratified 2D turbulence~\citep{McW1984, McW1990, Benzi1987, Benzi1988, Carnevale1991, Dritschel2008}.

With one exception, the studies in tables~\ref{Table1} and~\ref{Table2} report that $p=1/5$ is the first power-law scaling observed as $\Ra$ is increased from small values with fixed $\Pr$ (even with $\Pr \leq 1$).
Based on \citet{tsaiSheard20}, and our results in section~\ref{RaNu}, one  might speculate  that the first exponent  $p=1/5$ is, at sufficiently high $\Ra$, replaced by $p=1/4$.
The exception, however, is the ``complex scaling dependence'' reported by \cite{ReiterShish2020}.
With $\Pr = 1$ and $10$, and at relatively low $\Ra$, \cite{ReiterShish2020} report first a range with $p=1/4$.
As $\Ra$ is increased  there is a transition to $p=1/5$, followed at even higher $\Ra$ by second transition back to $p\approx 1/4$.
The situation is further complicated because with $\Pr = 0.1$ \cite{ReiterShish2020} find  only a single scaling regime with $p=1/5$.

Three-dimensionality distinguishes \cite{ReiterShish2020} from the 2D studies in table~\ref{Table1}.
But the comparable 3D study by \cite{G14} finds only $p=1/5$ regimes.
In the 3D solutions reported below we also find only  $p=1/5$ regimes.
The computations of \cite{ReiterShish2020} reach the highest values of $\Ra$ in 3D HC reported in the literature.
But the differences summarized above are at moderate values of $\Ra$ where there are several decades of overlap with other studies of HC (including this one).
The most striking difference of \cite{ReiterShish2020} from other work summarized in tables~\ref{Table1} and~\ref{Table2} is that with $\Pr = 1$ and $10$, the exponent $p=1/4$ is the first exponent encountered as $\Ra$ is increased from small values.
The piecewise constant buoyancy profile used as a surface boundary condition might be responsible for the distinctive $\Ra$--$\Nu$ scaling reported by \cite{ReiterShish2020}.

 In support of the hypothesis that the surface boundary condition might affect $p$,  the  2D $\Pr = \infty$ study of \cite{RH2019}  shows  a transition from steady  to unsteady flow at $\Ra \approx 10^8$.
 In the case of a piecewise constant buoyancy profile  (but not with a linear surface buoyancy profile) this transition is accompanied by an increase in $p$ from $1/5$ to $\approx 1/4$.
 Note, however, that consistent of \cite{R65}, and with \cite{CWHL08}, \cite{RH2019}  do  find that $p=1/5$ as the first  exponent encountered as $\Ra$ is increased from small values.

In section~\ref{form} we formulate the horizontal convection problem.
Section~\ref{RaNu} is a numerical study of HC in which we vary $\Ra$ at  $\Pr = 1$.
All solutions use the sinusoidal surface buoyancy profile in~\eqref{sbuoy8}.
There are four solution suites corresponding to   no-slip and free-slip  boundary conditions and to  2D and 3D solutions.
Section~\ref{RossbySec} is a  review of various $\Ra$--$\Nu$ scaling arguments leading  to $p=1/5$.
Using the  mechanical energy power-integral of HC  we put these arguments into a unified framework.
Section~\ref{RossbySec} also   shows   that the $p=1/4$ scaling regime is characterized by a ``double  boundary layer'':  there is a thin  BL, with thickness  $\sim \Ra^{-1/4}$, nested   inside a wider BL with thickness $\sim \Ra^{-1/5}$.
Section~\ref{oneQuarter} provides further characterization of the double BL structure of the $p=1/4$ scaling regime.
Section~\ref{2DT} shows that the exponent $p=1/4$  follows from the assumption that the thickness of  the inner BL is determined by the Kolmogorov and Batchelor length scales of HC.
In  section~\ref{3Dffs} we show that vorticity amplification by strain mediated vortex stretching does not operate in 3D HC: as $\Ra \to \infty$ the mean-square vorticity in 2D and 3D solutions are almost the same.
Section~\ref{conclusion2} concludes.

\section{Formulation of the horizontal convection problem \label{form}}

We consider a Boussinesq fluid with density $\rho = \rho_0 (1 - g^{-1} b)$, where $\rho_0$ is a constant reference density and $b$ is the ``buoyancy''.
If, for example, the fluid is stratified by temperature variations then $b = g \alpha (T - T_0)$, where $T_0$ is a reference temperature and $\alpha$ is the thermal expansion coefficient.
The Boussinesq equations of motion are
\begin{align}
  \bu_t + \bu \bcdot \grad \bu + \bnabla p &= b \boldsymbol{\hat z} + \nu \lap \bu \com \label{mom} \\
  b_t + \bu \bcdot \grad b &= \kappa \lap b \com \label{buoy} \\
  \bnabla \bcdot \bu &= 0 \per \label{divu}
\end{align}
The kinematic viscosity is $\nu$ and the thermal diffusivity is $\kappa$.

\subsection{Horizontal convective boundary conditions and control parameters}

We suppose the fluid occupies a domain with depth $h$, length $\ellx$, width~$\elly$; we assume periodicity in the $x$- and $y$-directions.
At the bottom surface ($z = 0$) and top surface ($z = h)$ the primary boundary conditions on the velocity, $\bu = (u, v, w)$, is that $w = 0$; the viscous boundary condition is either no slip (\NS{}), $u = v = 0$, or free slip (\FS{}), $u_z = v_z = 0$.
At the bottom $z = 0$ the buoyancy boundary condition is no flux, $\kappa b_z = 0$.
At the top, $z = h$, the boundary condition is $b = \bs(x)$, where the top surface buoyancy $\bs$ is a prescribed function of~$x$.
As a surface buoyancy field we use
\beq
  \bs(x) = \bstar \cos k x \com
  \label{sbuoy8}
\eeq
where $k = 2 \pi / \ellx$.

As an idealization of conditions at the sea surface, \FS{} is better than \NS.
But the main reason for considering different viscous boundary conditions is to test scaling arguments.
We find only minor quantitative differences in the $\Nu$--$\Ra$ scaling between the two boundary conditions.
In the numerical solutions described below the main features of the $\Nu$--$\Ra$ scaling relation are independent of the viscous boundary condition.

The problem is characterized by four non-dimensional parameters: the Rayleigh and Prandtl numbers
\beq
  \Ra \defn \frac{\ellx^3 \bstar}{\nu \kappa} \com \qquad \text{and} \qquad \Pr \defn \frac{\nu}{\kappa} \com
  \label{controlDef1}
\eeq
and the aspect ratios $\ellx / h$ and $\elly / h$.
With periodic boundary conditions in $y$ (no side walls), 2D HC is the special case $\elly / h = 0$.
The Rayleigh number in~\eqref{controlDef1} is defined using the amplitude $\bstar$ of the sinusoid in~\eqref{sbuoy8}.
Other authors use the total difference (peak-to-trough) in buoyancy to define $\Ra$.
For comparison one might multiply our numerical values of $\Ra$ in section~\ref{RaNu} by two.

\subsection{Mechanical energy dissipation}

We use an overbar to denote an average over $x$, $y$, and~$t$, taken at any fixed $z$ and angle brackets $\la \; \ra$ to denote a total volume average over $x$, $y$, $z$, and~$t$.
Using this notation, we recall some results from \cite{PY02} that are used below.

Horizontally averaging the buoyancy equation~\eqref{buoy} we obtain the zero-flux constraint
\beq
  \overline{wb} - \kappa \bar b_z = 0 \per
  \label{PY1}
\eeq
Taking $\la \bu \bcdot \eqref{mom} \ra$, we obtain the kinetic energy power integral
\begin{align}
  \varepsilon = \la w b \ra \com \label{PY2}
\end{align}
where $\varepsilon \defn \nu \la |\grad \bu|^2 \ra$ is the rate of dissipation of kinetic energy and $\la w b \ra$ is rate of conversion between potential and kinetic energy.

Vertically integrating~\eqref{PY1} from $z = 0$ to $h$, and using $\overline{\bs(x)} = 0$, we obtain another expression for $\la wb \ra$.  Substituting this into~\eqref{PY2} 
\begin{align}
  \varepsilon = - \frac{\kappa \bar b(0)}{h} \per \label{PY2.5}
\end{align}
In~\eqref{PY2.5}, $\bar b(0)$ is the $(x, y, t)$-average of the buoyancy at the bottom $z = 0$.

\subsection{The Nusselt number of horizontal convection}

Following \cite{RocNussDef}, we use the dissipation of buoyancy variance,
\beq
  \chi \defn \kappa \la |\bnabla b|^2 \ra \com
  \label{chidef}
\eeq
to define the Nusselt number as
\beq
  \Nu \defn {\chi} \big\slash{\chidiff} \per
  \label{Sig}
\eeq
Above, $\chidiff \defn \kappa \la |\grad \bdiff|^2 \ra$ is the buoyancy dissipation of the diffusive solution, i.e., $\kappa \lap \bdiff = 0$ with $\bdiff$ satisfying the same boundary conditions as $b$.

Application of variational methods to HC \citep{SKB04, WY09, RBLSY} results in bounds on $\chi$ taking the form $\Nu \lesssim \Ra^{1/3}$.
The exponent $1/3$ is safely larger than the exponents $1/5$ and $1/4$ reported in numerical studies of HC, including this work.

\cite{RocNussDef} show that there is also a ``surface Nusselt number''
\beq
  \Nus \defn \overline{\bs \, \kappa b_z(h)}\, \Big \slash \, \overline{\bs \, \kappa {\bdiff}_z(h)} \per
  \label{surfNussl}
\eeq
Above, $\kappa b_z(h)$ is the buoyancy flux through the top surface $z = h$.
With sufficient temporal averaging $\Nu = \Nus$. The interior entropy production, $\chi$, is balanced by an entropy flux through the surface $z = h$.
$\Nus$ is the non-dimensional entropy flux though the surface.
In numerical solutions described below, in which the temporal average is computed over a finite time interval, $\Nu \approx \Nus$ is a check on the estimated Nusselt number.


\def\dag{\dagger}
 \begin{table}
 \centering
\begin{tabular}{ c c c c c c}
 & \multicolumn{2}{c}{{Free-slip} $\Nu$} & \multicolumn{2}{c}{{No-slip} $\Nu$} & Highest resolution \\
 $\Ra$ & 2D$\,\textcolor{blue}{\circ}$ & 3D$\,\textcolor{blue}{\bullet}$ & 2D$\,\textcolor{red}{\ssquare}$& 3D$\,\textcolor{red}{\sbsquare}$ & $n_x$, $n_z$ \\ \hline
1.28e03 &   1.00 	   & 	1.00$^\dag$	 & 	 1.00 		&  1.00$^\dag$      & 128, 32 \\
3.20e03 &   1.02 	   &  1.02$^\dag$	 & 	 1.00	  	&  1.00$^\dag$      & 128, 32 \\
4.48e03 &   1.04 	   &	1.04$^\dag$	 &	 1.00	  	&  1.00$^\dag$      & 128, 32 \\
6.40e03 &   1.08 	   & 	1.08$^\dag$	 &   1.01 	  &  1.01$^\dag$      & 128, 32 \\
1.28e04 &   1.22 	   &	1.22$^\dag$	 &   1.02 	  &  1.02$^\dag$      & 128, 32 \\
1.92e04 &  	1.38 	   &	1.38$^\dag$	 &   1.04 	  &  1.04$^\dag$      & 128, 32 \\
3.20e04 &   1.64 	   & 	1.64$^\dag$	 &   1.11 	  &  1.11$^\dag$      & 128, 32 \\
6.40e04 &   2.07 	   &  2.07$^\dag$	 &   1.28 	  &  1.28$^\dag$      & 128, 32 \\
1.28e05 &  	2.55 	   &  2.55$^\dag$	 &   1.59 	  &  1.59$^\dag$      & 128, 32 \\
2.56e05 &   3.04 	   &	3.04$^\dag$	 &   1.96 	  &  1.96$^\dag$      & 128, 32 \\
6.40e05 &   3.71 	   & 	3.71$^\dag$	 &   2.47 	  &  2.47$^\dag$      & 128, 32 \\
1.60e06 &   4.50     &  4.50$^\dag$  &   3.01 	  &  3.01$^\dag$      & 256, 64 \\
3.20e06 &   5.14     &  5.14$^\dag$  &   3.45 	  &  3.45$^\dag$      & 256, 64 \\
6.40e06 &   5.80     &  5.80$^\dag$  &   3.93 	  &  3.93$^\dag$      & 256, 64 \\
1.60e07 &   6.77     &  6.77$^\dag$  &   4.70 	  &  4.70$^\dag$      & 256, 64 \\
3.20e07 &   7.61     &  7.61$^\dag$  &   5.38 	  &  5.44$^{\sharp}$  & 256, 64 \\
6.40e07 &   8.65     &  8.68$^*$ 	   &   6.17 	  &  6.59$^{\sharp}$	& 256, 64 \\
1.60e08 &  10.48     & 10.57$^*$ 	   &   7.41 	  &  8.46$^{\sharp}$  & 256, 64 \\
3.20e08 &  12.17$^*$ & 12.19$^*$ 	   &   8.51 	  & 10.11$^{\sharp}$ 	& 256, 64 \\
6.40e08 &  14.01$^*$ & 14.09$^*$     &   9.69 	  & 11.97$^*$ 	      & 256, 64 \\
1.60e09 &  16.99$^*$ & 17.07$^*$ 		 &  11.77$^*$ & 14.75$^*$ 	      & 256, 64 \\
3.20e09 &  19.60$^*$ & 20.32$^*$ 		 &  13.53$^*$ & 17.20$^*$ 	      & 512, 128 \\
6.40e09 &  22.41$^*$ & 24.64$^*$ 		 &  15.35$^*$ & 19.86$^*$ 	      & 512, 128 \\
1.60e10 &  27.32$^*$ & 31.38$^*$ 		 &  18.83$^*$ & 23.67$^*$ 	      & 512, 128 \\
3.20e10 &  31.48$^*$ & 37.28$^*$ 		 &  21.88$^*$ & 27.61$^*$ 	      & 512, 128 \\
6.40e10 &  36.08$^*$ & 43.94$^*$ 		 &  25.50$^*$ & 30.68$^*$ 	      & 512, 128 \\
1.28e11 &  41.85$^*$ & 49.74$^*$ 		 &  29.87$^*$ & 36.67$^*$ 	      & 1024, 256 \\
1.60e11 &  43.80$^*$ &           		 &  32.02$^*$ &                   & 1024, 256 \\
3.20e11 &  50.71$^*$ & 59.91$^*$     &  37.45$^*$ & 44.01$^*$         & 1024, 256 \\
6.40e11 &  58.36$^*$ & 				       &  44.23$^*$ &                   & 1024, 256 \\
1.60e12 &  71.43$^*$ & 				       &  55.65$^*$ &                   & 1024, 256 \\
3.20e12 &  86.26$^*$ & 				       &  66.97$^*$ &                   & 2048, 512 \\
6.40e12 & 101.32$^*$ & 				       &  78.86$^*$ &                   & 4096, 1024 \\
1.60e13 & 128.44$^*$ & 				       &  97.66$^*$ &                   & 4096, 1024 \\
6.40e13 & 178.55$^*$ & 				       & 133.16$^*$ &                   & 4096, 1024 \\
\end{tabular}
\caption{ $\Nu$--$\Ra$ data for HC DNS. All runs have $Pr = 1$ and $\ellx / h = 4$. 3D runs have $\elly / h = 1$ and $n_y = n_z$. The surface buoyancy is the sinusoid in~\eqref{sbuoy8}. Unsteady solutions are indicated by a superscript $*$ on $\Nu$; strictly 2D solutions (no $y$-dependence and $v = 0$) of 3D computations are marked by a superscript $\dagger$. The four \NS{} runs with superscript $\sharp$ are 3D but steady.} \label{summarytable}
\end{table}

\section{A numerical study of horizontal convection with $\Pr = 1$ \label{RaNu}}

In this section we present the results of a numerical study directed at characterizing the variation of the Nusselt number $\Nu$ in~\eqref{Sig} as a function of $\Ra$.
Computations are performed using Dedalus, a spectral framework for solving partial differential equations \citep[][\url{www.dedalus-project.org}]{Dedalus2020}.
We use Fourier bases in the horizontal, periodic directions and a Chebyshev basis in the vertical; the equations are time stepped using a fourth-order implicit-explicit Runge--Kutta scheme.

\begin{figure}
\centering
  \includegraphics[width = 1\textwidth]{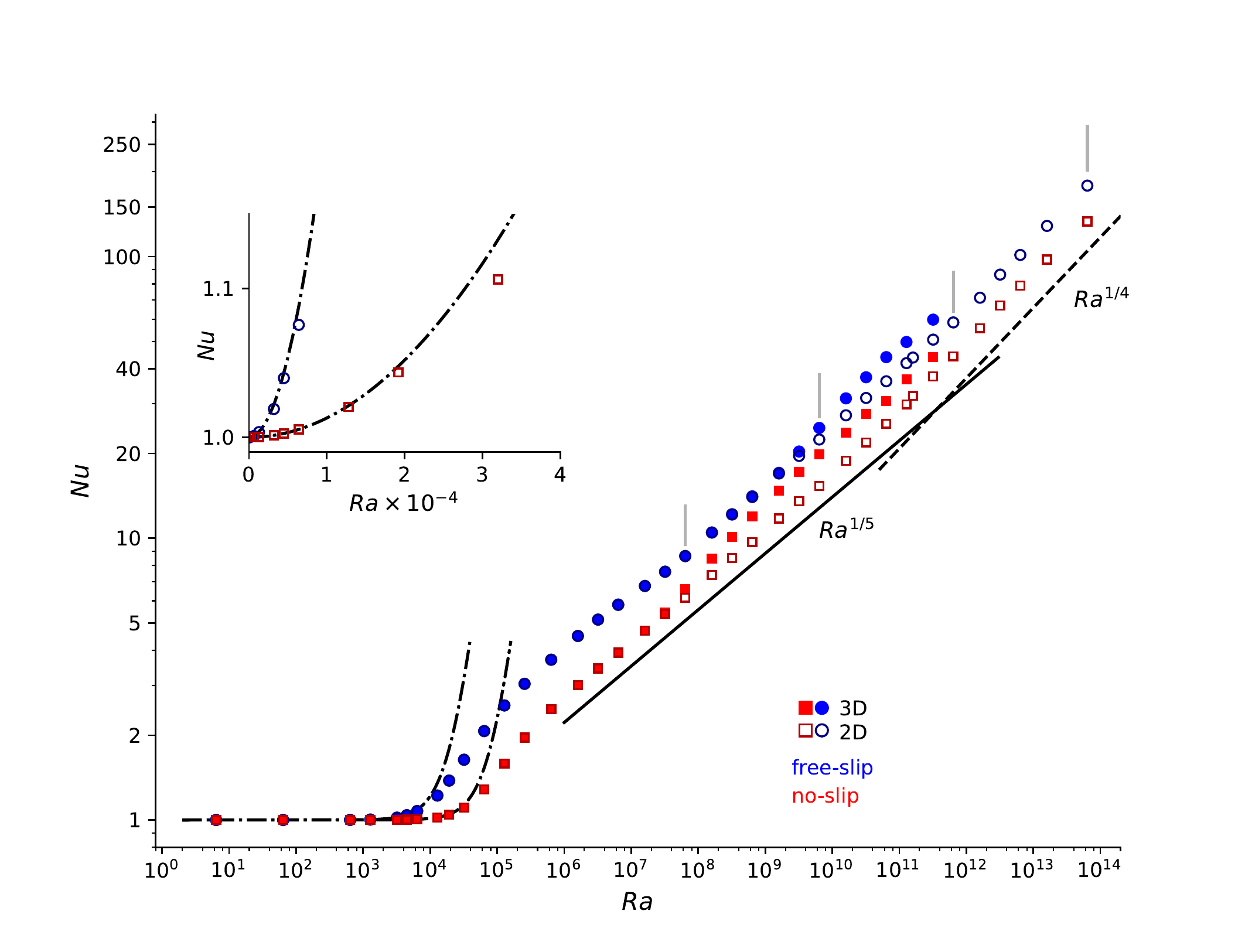}
  \caption{Variation of Nusselt number $\Nu$ with Rayleigh number $\Ra$ using the data from table~\ref{summarytable}. The inset compares the low-$\Ra$ numerical results with the low-$\Ra$ analytic results~\eqref{nstud3} and~\eqref{nstud4}. Some solid markers fall on top of open markers, indicating that the 3D solutions evolve to become 2D, or that the three dimensionality is weak. The four vertical grey line segments mark $\Ra$'s of the solutions in figures~\ref{Fig1} and~\ref{Fig2}.}
 \label{Fig3}
\end{figure}

We limit attention to $\Pr = 1$ and the sinusoidal surface buoyancy forcing~$\bs(x)$ in~\eqref{sbuoy8}.
We discuss both \NS{} and \FS{} boundary conditions and consider 2D solutions with aspect ratios
\beq
  \ellx / h= 4 \com \qquad \elly / h = 0 \com
\eeq
and 3D solutions with
\beq
  \ellx / h = 4 \com \qquad \elly / h = 1 \per
\eeq
Thus we have four solution suites: \twoDFS, \threeDFS, \twoDNS, and \threeDNS.
The  estimates of $\Nu$ are summarized in table~\ref{summarytable} and figure~\ref{Fig3}.

\subsection{The low-$\Ra$ regime}

Analysis of the low-$\Ra$ regime in appendix~\ref{lowRa} shows that with $\ellx / h = 4$ the first variation of the Nusselt number away from unity is
\beq
  \Nu_{}^{\!\!\!\! \FS} = 1 + \left( \frac{\Ra}{21\, 567.5} \right)^2 + \text{ord}\big(\Ra^4\big) \com \label{nstud3}
\eeq
and
\beq
  \Nu^{\!\!\!\! \NS} = 1 + \left( \frac{\Ra}{87\, 789.8} \right)^2 + \text{ord}\big(\Ra^4\big) \per
\label{nstud4}
\eeq
The low-$\Ra$ regime means that the $\Ra^2$ term in~\eqref{nstud3} and~\eqref{nstud4} is less than one, i.e., that the convective buoyancy transport is a weak enhancement of the diffusive transport.
For \FS{} low $\Ra$ means that $\Ra$ is somewhat less than about $2.5 \times 10^3$ and for \NS{} low $\Ra$ means that $\Ra$ is somewhat less than about $10^4$.
These analytic results are compared with numerical solutions in the insert of figure~\ref{Fig3}.
For example, the \twoDFS{} solution at $\Ra = 6.4$ has $\Nu - 1 = 8.8 \times 10^{-8}$ and at $\Ra = 640$, $\Nu - 1 = 8.8 \times 10^{-4}$.
The approximations in~\eqref{nstud3} and~\eqref{nstud4}  are very accurate for $\Ra < 10^3$. Table~\ref{summarytable} does not include  these very low-$\Ra$ results.

\subsection{$\Nu$--$\Ra$ scaling regimes: $p=1/5$ and $p=1/4$}

\begin{figure}
  \centering
  \includegraphics[width=0.9\textwidth]{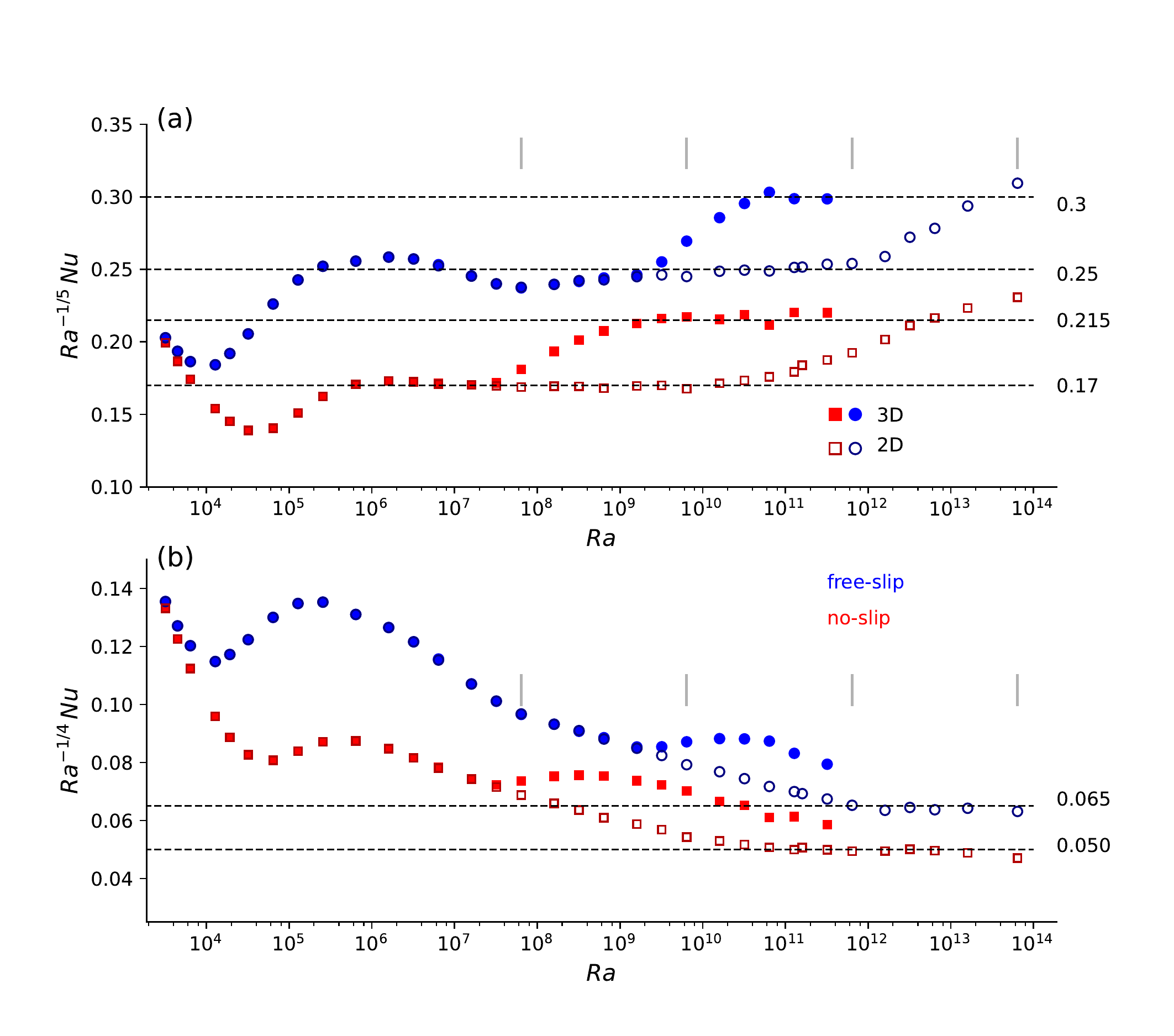}
  \caption{Variation of ``compensated Nusselt numbers'' (a)~$\Ra^{-1/5} \Nu$ and (b)~$\Ra^{-1/4} \Nu$ with Rayleigh number $\Ra$. The four vertical grey line segments mark $\Ra$'s of the solutions in figures~\ref{Fig1} and~\ref{Fig2}.}
  \label{Fig4}
\end{figure}

Between $\Ra \sim 10^4$ and $10^5$ we do not see a simple relation between $\Nu$ and $\Ra$.
But once $\Ra$ is greater than about $6.4 \times 10^5$ we find the $p=1/5$ scaling,
\beq
  \Nu \sim \Kfive \Ra^{1/5} \com \label{RossbyScaling}
\eeq
in all four solution suites.
Starting at around $\Ra \sim 10^{11}$ in the \twoDNS{} suite, and $10^{12}$ in the \twoDFS{} suite, there is a transition from the regime~\eqref{RossbyScaling} to the $p=1/4$ scaling
\beq
  \Nu \sim \Kfour \, \Ra^{1/4} \per \label{oneFourth}
\eeq
The $p=1/4$ regime with \NS{} has been reported  previously by \citet{tsaiSheard20}.

 
In figure~\ref{Fig4} we show the data from figure~\ref{Fig3}, replotted using the compensated Nusselt number $\Ra^{-1/5} \Nu$ in panel~(a) and $\Ra^{-1/4} \Nu$ in panel~(b).
Table~\ref{leastSquares1} summarizes the exponents determined by least-squares fitting the $\Ra$--$\Nu$ data over selected ranges. Least-squares exponents are broadly in agreement with the scaling regimes determined by visual inspection of figure~\ref{Fig4} and other compensated plots.
We use least-squares because it is objective and reproducible.
Least squares also assesses the sensitivity of estimated exponents to the points at the beginning and end of a putative scaling range.
 

\subsection{Discussion of the no-slip solutions}

We begin with the \twoDNS{} solution suite.
The $p=1/5$ scaling~\eqref{RossbyScaling} is found across the four-decade range in row~1 of table~\ref{leastSquares1}. This is the plateau at $\Kfive = 0.17$ in figure~\ref{Fig4}(a).
Least-squares estimates of exponent and prefactor, $\Kfive$, are robust to changes in the range, e.g., row~2 of table~\ref{leastSquares1}.
The \twoDNS{} suite transitions to the $p=1/4$ scaling~\eqref{oneFourth} at around $\Ra = 10^{11}$ and forms the plateau at $\Kfour = 0.05$ in figure~\ref{Fig4}(b); see rows~4 through~6 of table~\ref{leastSquares1}.


The \threeDNS{} solution suite is more complicated.
With moderate $\Ra$ (rows~7 through~9 of table~\ref{leastSquares1}) the \threeDNS{} solutions coincide with their \twoDNS{} partners and the scaling is again~\eqref{RossbyScaling} with $\Kfive = 0.17$.
At a critical $\Ra$ of roughly $3.2 \times 10^7$ the \threeDNS{} suite becomes unstable to 3D perturbations.
With further increases in $\Ra$ the \threeDNS{} solutions have larger $\Nu$ than their \twoDNS{} colleagues: the four steady \threeDNS{} solutions in the interval $3.20\times 10^{7} \leq \Ra \leq 3.20\times 10^8$, are here.
One might expect that development of 3D flow, albeit steady 3D flow, signals the beginning of a new scaling regime, with an exponent $p$ greater than $1/5$.
But alas, this is the transition discovered by~\cite{G14}: at about $\Ra = 1.60\times 10^9$ the \threeDNS{} solutions enter a new $p=1/5$ regime: see rows~10 through~12 of table~\ref{leastSquares1} and the \threeDNS{} plateau at $0.215$ in figure~\ref{Fig4}(a).
With maximum $\Ra = 3.20 \times 10^{11}$, we did not find convincing evidence of $p=1/4$ in the \threeDNS{} solution suite.
 
The \NS{} computations of \cite{G14} used $\Pr=5$ and a piecewise constant surface buoyancy.
Instead of~\eqref{Sig}, \cite{G14} defined $\Nu$ based on the buoyancy flux through the destabilized portion of the non-uniformly heated surface.
Despite these differences, \cite{G14} document analogous \threeDNS{} behavior involving  two $p=1/5$ scaling regimes: the constant $\Kfive$ takes different values on either side of a smooth transition.
In the second $p=1/5$ regime the results of \cite{G14} are based on large eddy simulation (LES). But our direct numerical solutions also reenter the $p=1/5$ regime.
Reentry to $p=1/5$ cannot be dismissed as an artifact of LES.

\begin{table}
  \centering
  \begin{tabular}{ c| c | c | c | c}
  row & suite &  range  &         points     & least squares  $\Nu$ \\
  \hline
   1 & \twoDNS    & $6.40 \times 10^{ 5} \leq \Ra \leq 6.40 \times 10^{ 9}$ & 13  & $0.177 \, \Ra^{0.198}$\\
   2 & \twoDNS    & $1.60 \times 10^{ 6} \leq \Ra \leq 3.20 \times 10^{ 9}$ & 11  & $0.178 \, \Ra^{0.197}$ \\
  {} & {}         & {} & {} & \\
   4 & \twoDNS    & $6.40 \times 10^{10} \leq \Ra \leq 6.40 \times 10^{13}$ & 10  & $0.062 \, \Ra^{0.242}$ \\
   5 & \twoDNS    & $6.40 \times 10^{10} \leq \Ra \leq 1.60 \times 10^{13}$ &  9  & $0.056 \, \Ra^{0.246}$ \\
   6 & \twoDNS    & $1.28 \times 10^{11} \leq \Ra \leq 1.60 \times 10^{13}$ &  8  & $0.055 \, \Ra^{0.246}$ \\
  {} & {}         & {} & {} & \\
   7 & \threeDNS  & $6.40 \times 10^{ 5} \leq \Ra \leq 3.20 \times 10^{ 7}$ &  6  & $0.173 \, \Ra^{0.199}$ \\
   8 & \threeDNS  & $6.40 \times 10^{ 5} \leq \Ra \leq 6.40 \times 10^{ 6}$ &  4  & $0.167 \, \Ra^{0.202}$ \\
   9 & \threeDNS  & $3.20 \times 10^{ 6} \leq \Ra \leq 3.20 \times 10^{ 7}$ &  4  & $0.179 \, \Ra^{0.197}$ \\
  {} & {}         & {} & {} & \\
  10 & \threeDNS  & $1.60 \times 10^{ 9} \leq \Ra \leq 3.20 \times 10^{11}$ &  8  & $0.195 \, \Ra^{0.204}$ \\
  11 & \threeDNS  & $1.60 \times 10^{ 9} \leq \Ra \leq 1.60 \times 10^{10}$ &  4  & $0.191 \, \Ra^{0.205}$ \\
  12 & \threeDNS  & $3.20 \times 10^{10} \leq \Ra \leq 3.20 \times 10^{11}$ &  4  & $0.180 \, \Ra^{0.208}$ \\
  {} & {}         & {} & {} & \\
  20 & \twoDFS    & $6.40 \times 10^{ 5} \leq \Ra \leq 1.60 \times 10^{11}$ & 18  & $0.253 \, \Ra^{0.199}$ \\
  21 & \twoDFS    & $6.40 \times 10^{ 5} \leq \Ra \leq 3.20 \times 10^{ 8}$ &  9  & $0.312 \, \Ra^{0.186}$ \\
  22 & \twoDFS    & $6.40 \times 10^{ 8} \leq \Ra \leq 1.60 \times 10^{11}$ &  9  & $0.215 \, \Ra^{0.206}$ \\
  {} & {}         & {} & {} & \\
  23 & \twoDFS    & $6.40 \times 10^{11} \leq \Ra \leq 6.40 \times 10^{13}$ &  6  & $0.074 \, \Ra^{0.245}$ \\
  24 & \twoDFS    & $6.40 \times 10^{11} \leq \Ra \leq 3.20 \times 10^{12}$ &  3  & $0.082 \, \Ra^{0.241}$ \\
  25 & \twoDFS    & $6.40 \times 10^{12} \leq \Ra \leq 6.40 \times 10^{13}$ &  3  & $0.073 \, \Ra^{0.245}$ \\
  {} & {}         & {} & {} & \\
  26 & \threeDFS  & $6.40 \times 10^{ 5} \leq \Ra \leq 3.20 \times 10^{ 8}$ &  9  & $0.308 \, \Ra^{0.187}$ \\
  {} & {}         & {} & {} & \\
  27 & \threeDFS  & $6.40 \times 10^{10} \leq \Ra \leq 3.20 \times 10^{11}$ &  3  & $0.358 \, \Ra^{0.193}$ \\
  \end{tabular}
  \caption{Summary of least-squares fits to various scaling regimes. Where possible, we assess the sensitivity of the exponent by varying the range.} \label{leastSquares1}
\end{table}

\subsection{Discussion of the free-slip solutions}
 
Turning to the \twoDFS{} solutions, the most generous identification of the $p=1/5$ regime in~\eqref{RossbyScaling} is the five-decade range in row 20 of table~\ref{leastSquares1}.
These 18 points correspond to the plateau $\Kfive = 0.25$ in figure~\ref{Fig4}(a).
We are concerned, however, by 9 points in the first half of this range, i.e., row 21 of table~\ref{leastSquares1}.
These 9 points undulate around the $\Kfive=0.25$ plateau with an amplitude of about $\pm 0.01$ and the least-squares exponent $0.186$ is uncomfortably different from $1/5$.
These  wayward points, at only moderately large $\Ra$, correspond to solutions that are either steady, or weakly time dependent.
Thus insufficient time-averaging in the estimate of $\Nu$ is not an issue.
Moreover in this range the 2D and 3D solutions coincide.
We conducted several tests by changing the spatial resolution and found no significant variation in the numerical estimate of $\Nu$.
If one views the exponent $0.186$ as close to $1/5$ then the wayward points are the lower end of a five-decade \twoDFS{} scaling regime: the undulation is a pre-asymptotic imperfection in the first half of this regime.
A more cautious interpretation is that the \twoDFS{}  $p=1/5$ regime begins only at about $\Ra = 6.40 \times 10^8$ and consists of the 9 points in row~22 of table~\ref{leastSquares1}. The \twoDFS{} suite transitions to $p=1/4$ scaling~\eqref{oneFourth} at around $\Ra = 10^{12}$ and forms the plateau at $\Kfour=0.065$ in figure~\ref{Fig4}(b); see rows 23 through 25 of table~\ref{leastSquares1}.

The \threeDFS{} solutions depart significantly from their \twoDFS{} colleagues first at about $\Ra = 3.20 \times 10^9$. There is no  evidence of $p=1/4$ scaling in the \threeDFS{} suite. Instead, the three  \threeDFS{} solutions in row 27 of table~\ref{leastSquares1} indicate a second  $p=1/5$ regime e.g. the $0.3$ plateau in figure~\ref{Fig4}(a). We speculate that the  \threeDFS{} suite     is recapitulating the phenomenology seen in the \threeDNS{} suite:  two $p=1/5$ scaling regimes separated by a smooth transition. This speculation is based  on the three solutions in row 27 which span less than one decade of variation in $\Ra$.

\section{Review of $\Nu \sim \Ra^{1/5}$ scaling arguments \label{RossbySec}}

\cite{R65} proposed a visco-diffusive balance in the boundary layer adjacent to the non-uniformly heated surface and so arrived at the one-fifth scaling in~\eqref{RossbyScaling}.
Rossby identified the length scale
\beq
  \dafive \defn \Ra^{-1/5}h
  \label{deltaRo}
\eeq
as the thickness of the surface BL.
In the following discussion we also need the length 
\beq
  \dafour \defn \Ra^{-1/4}h \per
  \label{dafourdef}
\eeq 
At $\Ra = 6.4 \times 10^{13}$, the ratio of these two BL scales is $\dafive / \dafour \approx 5$. 
 
Central to Rossby's $\Pr \gg 1$ argument is the assumption that BL buoyancy forces are balanced by viscosity and that BL inertia is subdominant.
At moderately large $\Ra$, the exponent $1/5$ has been supported by subsequent laboratory work and by numerical studies \citep{R98, M04, SKB04, WH2005, CWHL08, SK2011, IV12}.

 

\begin{figure}
  \centering
  \includegraphics[width=0.9\textwidth]{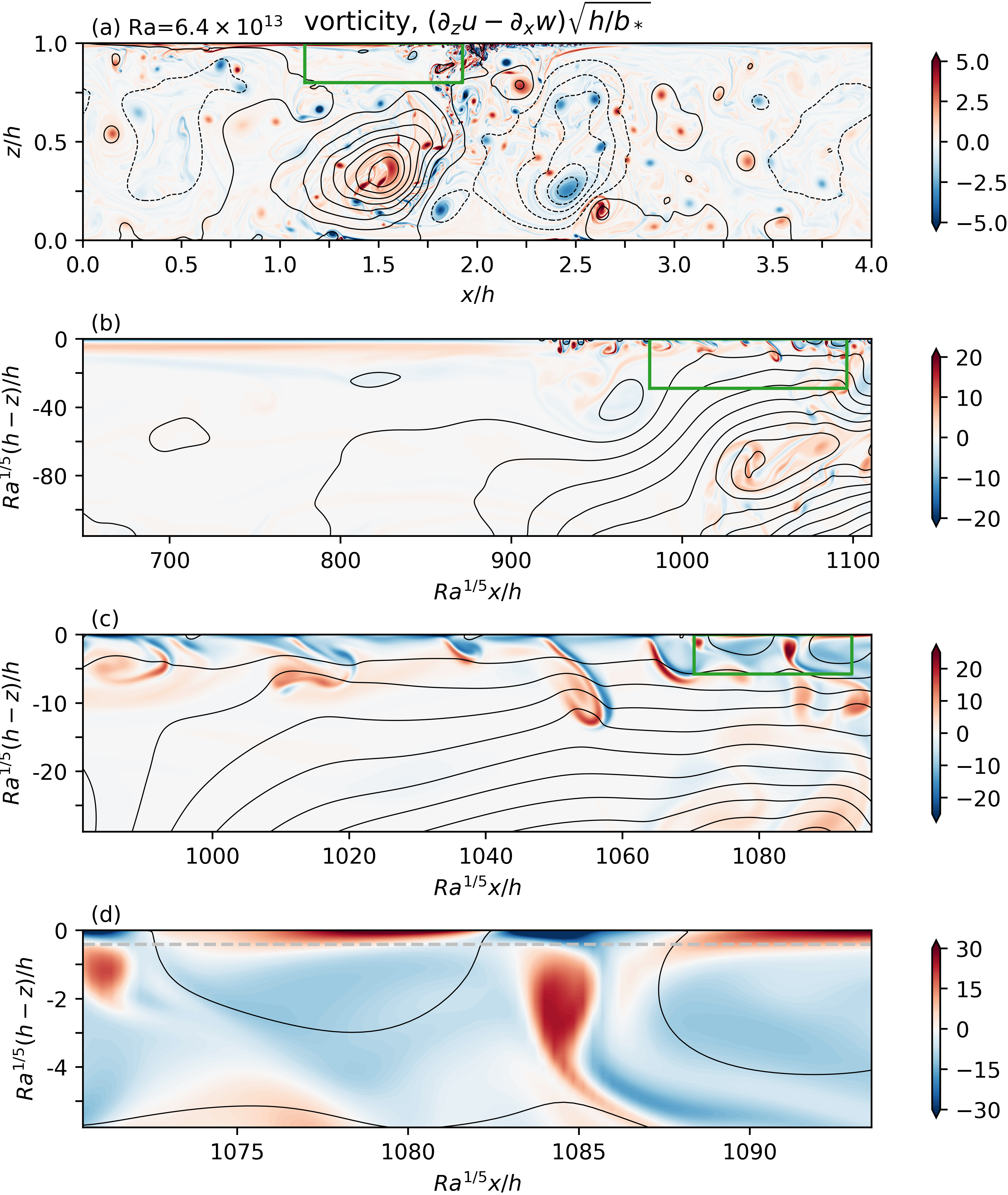}
  \caption{(a) A snapshot of vorticity in the \twoDNS{} solution at $\Ra = 6.4 \times 10^{13}$; this solution is in the non-Rossby $\Nu \sim \Ra^{1/4}$ scaling regime. Panels~(b)-(d) depict the boundary-layer structure by progressively zooming in to the top surface. Green rectangles in panels~(a), (b) and~(c) indicate the regions in panels~(b),~(c) and~(d) respectively. In panels~(b),~(c) and~(d), both axes are measured in units of $\dafive$. The dashed grey line in panel~(d) indicates the distance $2 \dafour$ below the top surface $z = h$. The contours in all panels are streamlines.}
  \label{Fig5}
\end{figure}

We emphasize that the $p=1/5$ scaling~\eqref{RossbyScaling}, and the associated BL thickness $\dafive$,  seems not, however, to require that $\Pr \gg 1$. The 2D solutions shown in figure~\ref{Fig1} -- including the unsteady solution in panels~(c) and~(d) -- are well within the $\Kfive = 0.25$ regime of figure~\ref{Fig4}(a). Our  unsteady \twoDFS{} solutions exhibit the one-fifth scaling~\eqref{RossbyScaling} over at least three decades of $\Ra$. All these $p=1/5$ solutions have $\Pr=1$.


To further complicate the situation, 2D solutions in the $p=1/4$ regime~\eqref{oneFourth} still express the BL scale $\dafive$.  Figure~\ref{Fig5} shows a progressively expanded view of the structure of HC near the upper surface.
This \twoDNS{} solution is in the \textit{non}-Rossby $p=1/4$ scaling regime~\eqref{oneFourth}.
Nonetheless, panel~(d) of figure~\ref{Fig5} indicates that $\dafive$ is a useful BL length scale.
We conclude that at sufficiently high $\Ra$ there is a double BL:  a thin-$\dafour$ layer is nested within a thicker $\dafive$-layer.
We discuss this double BL further in section~\ref{oneQuarter}.

(The \twoDNS{} solution in figure~\ref{Fig5}(a) exhibits the vortex-gas phenomenology noted previously in the \twoDFS{} solutions shown in figure~\ref{Fig2}. At high $\Ra$, no matter the viscous boundary condition, the interior of 2D HC is characterized as a vortex gas.)

As an alternative to Rossby scaling, \cite{Shish16} proposed a set of scaling arguments summarized in a phase diagram of the $(\Ra, \Pr)$-plane.
This diagram shows high-$\Pr$ regions denoted $\mathrm{I}^*_\ell$, $\mathrm{I}_{\infty}$ and $\mathrm{III}_{\infty}$; these three high-$\Pr$ regions have $\Nu \sim \Ra^{p}$ with $p=1/4$ in $\mathrm{I}^*_\ell$ and $\mathrm{III}_{\infty}$ and $p=1/6$ in $\mathrm{I}_{\infty}$.
This tripartite proposal cannot be reconciled with the $\Pr\gg 1$ results of \cite{R65}, \cite{CWHL08} and \cite{RH2019}.
In the phase diagram of \cite{Shish16}, the $p=1/5$ scaling  is found only in the $\Pr\ll 1$ region $\mathrm{I}_{\ell}$.
We discuss region $\mathrm{I}_{\ell}$ in more detail below.

\subsection{The spanwise average}

To identify the various processes in the BL, we begin taking a spanwise $y$-average of the equations of motion.
Denote this spanwise average with a hat so that
\beq
  b(x, y, z, t) = \underbrace{\frac{1}{\elly} \int_0^{\elly} \! b(x, y, z, t) \, \dd y }_{\defn \hat{b}(x, z, t)} + \, b'(x, y, z, t) \per
  \label{span3}
\eeq
Above, $b'(x,y,z,t)$ is the three-dimensional departure from the spanwise average.
Taking the spanwise average of the 3D continuity equation~\eqref{divu}, we obtain a 2D ``overturning stream function'' $\psi(x, z, t)$, such that $(\hat u \, , \, \hat w) = (- \psi_z \, , \, \psi_x)$.
With this notation the 3D velocity is written as
\beq
  \left( u, v, w \right) = (- \psi_z \, , 0 , \psi_x) + (u', v', w') \per
  \label{overturning}
\eeq

The spanwise-average of the buoyancy equation is
\beq
\hat b_t + \psi_x \hat b_z - \psi_z \hat b_x + \p_x \, \widehat{u'b'} + \p_z \, \widehat{w'b'} = \kappa \lap \hat b \com
\label{buoy1}
\eeq
and the spanwise average of the spanwise vorticity equation is
\beq
  \underbrace{\zeta_t + \psi_x \zeta_z - \psi_z \zeta_x}_{\text{inertia}} \; + \!\! \underbrace{\hat b_x}_{\substack{\text{buoyancy} \\ \text{torque}} }
  \!\! + \; \underbrace{(\p_z^2 - \p_x^2) \, \widehat{u'w'} + \p_x\p_z \big(\widehat{u'^2} - \widehat{w'^2}\big) }_{\text{Reynolds stress torque (RST)}}
  = \underbrace{\nu \lap \zeta}_{\text{viscosity}} \!
\per
  \label{vorty7}
\eeq
Above $\zeta \defn - \lap \psi$ is the spanwise-averaged spanwise vorticity.
The power integral~\eqref{PY2.5} becomes
\begin{align}
  \varepsilon &= \nu \left \la \zeta^2 \right\ra + \nu \la |\grad \bu'|^2 \ra \label{powInt13} \\
              &= - \kappa \bar b(0) / h \per
  \label{powInt17}
\end{align}
The 2D equations of motion are recovered by suppressing the spanwise averages of quadratic fluctuations in~\eqref{buoy1}, \eqref{vorty7}, and~\eqref{powInt13} e.g. the 2D vorticity equation is~\eqref{vorty7} without RST.

\subsection{A review of $\Nu \sim \Ra^{1/5}$ scaling arguments \label{scaleArg}}

Following \cite{Shish16}, we assume that there is a BL with thickness $\dab$ in buoyancy and $\dau$ in momentum and vorticity.
The Reynolds number is
\beq
  \Re \defn \frac{U h}{\nu} \com
  \label{ReDef}
\eeq
where
\beq
  U \defn\sqrt{\la |\bu|^2 \ra} \label{UDef}
\eeq
is the typical flow velocity.
The domain length scales $\ellx$, $\elly$ and $h$ are roughly comparable. We use the depth $h$ as representative of the domain length scale.

Scale analysis of the surface Nusselt number in~\eqref{surfNussl}, e.g., $\hat b_z(h) \sim \bstar/ \dab$, shows that
\beq
  \Nu \sim \frac{\h}{\dab} \per \label{Nul}
\eeq
One reaches the same conclusion via scale analysis of the $\chi$-based Nusselt number in~\eqref{chidef}: although $|\grad b|^2 \sim \bstar^2 / \dab^2$, the $\chi$-BL occupies only a fraction $\dab / \h$ of the domain.
Thus~\eqref{Nul} follows because of the volume average $\la \; \ra$.

Now apply scale analysis to the buoyancy equation~\eqref{buoy1}.
Using results such as $\psi_z \hat b_x \sim \psi_x \hat b_z \sim U {\bstar}/\h$ and $\kappa \lap \hat b \sim \kappa \bstar/\dab^{2}$ one has
\beq
  U \sim \frac{\kappa \h}{\dab^{2}} \com \qquad \text{or in non-dimensional form} \qquad \Nu \sim (\Re \Pr)^{1/2} \per \label{everyone3}
\eeq

To estimate the viscous dissipation $\varepsilon$  on the right  of the power integral~\eqref{powInt13}, one assumes that an order-one fraction of $\varepsilon$ is concentrated in the BL, and this BL  occupies a fraction $\dau / \h$ of the domain.
One can either neglect $\nu \la |\grad \bu'|^2\ra$, or assume that both terms on the right of~\eqref{powInt13} scale in the same way, i.e.,~as $\nu \zeta^2 \sim \nu (U / \dau)^2$.
In either case 
\beq
  \varepsilon \sim \frac{\nu U^2}{\dau h} \per 
  \label{everyone7}
\eeq 
Scale analysis of the right hand side of~\eqref{powInt17} assumes that the bottom buoyancy, $\bar b(0)$, is an order-one fraction of the minimum buoyancy, $-\bstar$, on the top surface. (The stronger result  that $\bar b(0) \to - \bstar$ as $\Ra \to \infty$ is likely true.) Thus
\beq
  \varepsilon \sim \frac{\kappa \bstar}{h} \per
  \label{everyone8}
\eeq 
Combining~\eqref{everyone7} and~\eqref{everyone8} 
\beq
  U^2 \sim \frac{\kappa}{\nu}\dau\bstar\com \qquad \text{or in non-dimensional form} \qquad (\Re\Pr )^2 \sim \frac{\dau}{\h} \Ra \per
  \label{everyone11}
\eeq
Eliminating  $\Re \Pr$ between~\eqref{everyone3} and~\eqref{everyone11}, and then using~\eqref{Nul} to get rid of~$h$, one finds
 \beq
  \Nu^5 \sim \frac{\dau}{\dab} \Ra \per
  \label{everyone17}
\eeq
The final step to obtain the dependence of $\Nu$ on $\Ra$ and $\Pr$ is to express the ratio $\dau / \dab$ on the right of~\eqref{everyone17} in terms $\Ra$ and $\Pr$.
There are three arguments in the literature.

\textbf{The scaling of Rossby (1965).} Taking $\dau = \dab$ one obtains from~\eqref{everyone17}
\beq
  \Nu \sim \Pr^0\, \Ra^{1/5} \qquad \text{and} \qquad \Re \sim \Pr^{-1} \Ra^{2/5} \per
  \label{RossbyClassico}
\eeq
Rossby's 1965 argument did not employ the power integral and its consequence~\eqref{everyone17}.
Instead, Rossby assumes \textit{ab initio} that $\dab = \dau$ and balances buoyancy torque with viscosity in~\eqref{vorty7}, leading to $U \sim \bstar \dau^3 / \ell \nu$.
Combining these results with~\eqref{Nul} and~\eqref{everyone3} one again finds~\eqref{RossbyClassico}.
Rossby's balance between buoyancy torque and viscosity applies to both \FS{} and \NS.
In the \FS{} case, the velocity BL results from the vorticity source $\hat b_x$ in~\eqref{vorty7}: this rationalization of Rossby's assumption that $\dau = \dab$ also applies to \NS.
 
\textbf{The scaling of Gayen, Griffith \& Hughes (2014).} In the vorticity equation~\eqref{vorty7}, balance buoyancy torque with either inertia or Reynolds stress torques, leading to $U^2 \sim \dau \bstar$, and follow Rossby by assuming that $\dab = \dau$.
Combining these results with~\eqref{Nul} and~\eqref{everyone3} one finds
\beq
  \Nu \sim \Pr^{1/5} \Ra^{1/5} \qquad \text{and} \qquad \Re \sim \Pr^{-3/5} \Ra^{2/5}\per
  \label{GGH29}
\eeq
This argument does not use the power integral and it is not consistent with~\eqref{everyone17} unless $\Pr$ is order unity.
Using the scaling assumptions above to estimate $\varepsilon$ in ~\eqref{powInt13} and~\eqref{powInt17} we find
\beq
  \nu \laa \zeta^2 \raa \sim \Ra \, (\kappa \nu^2/\h^4) \qquad \text{and} \qquad -\kappa \bar b(0) /h \sim \Ra \, (\kappa^2 \nu/\h^4) \per
  \label{GGH31}
\eeq
The two terms in~\eqref{GGH31} differ by a factor of $\Pr$: this is a problem if $\Pr$ is either very large or very small.
But with $\Pr$ of order unity -- and here we consider $\Pr = 1$ -- there is no problem closing the mechanical energy budget and thus the scaling of \cite{G14} is a valid alternative to that of Rossby.

\textbf{The scaling of Shishkina, Grossman \& Lohse (2016).} In the vorticity equation~\eqref{vorty7}, balance inertia with viscosity, leading to $U\sim \nu \h / \dau^2$.
Eliminating $U$ with~\eqref{everyone3} one finds $\dau = \Pr^{1/2} \, \dab$, and substituting into the power-integral~\eqref{everyone17}
\beq
  \Nu \sim \Pr^{1/10} \, \Ra^{1/5}\com \qquad \text{and} \qquad \Re \sim \Pr ^{-4/5} \Ra^{2/5} \per
  \label{shishMess}
\eeq
A distinctive feature of this scaling argument is that buoyancy torque $\hat b_x$ in~\eqref{vorty7} does not appear in the leading-order BL vorticity balance.
This is justified by requiring that $\Pr\ll 1$, so that $\dau = \Pr^{1/2} \, \dab \ll \dab$.
In other words, this visco-inertial BL is so thin that both viscosity and inertia are much greater than the buoyancy torque $\hat b_x \sim \bstar/ h$.

Despite different physical assumptions, the three arguments summarized above are in agreement that $\Nu \sim \Ra^{1/5}$. All differences lie in the predicted dependence of $\Nu$ on~$\Pr$.
Moreover the three groups have presented numerical evidence in support of their particular $\Pr$-exponent \citep{R98,G14, SW16}.
Perhaps the different surface buoyancy profiles used by the three groups is important?
It is beyond our scope here to contribute further to investigation of the $\Pr$-exponent and the possible influence of the surface buoyancy profile.

We close this review by noting that  the nomenclature used to designate scaling~\eqref{shishMess} -- corresponding to region $\mathrm{I}_\ell$ in the phase diagram of \cite{Shish16} -- needs clarification.
Region $\mathrm{I}_\ell$, with $\Pr \ll 1$, is referred to by \cite{Shish16} as ``Rossby scaling''.
Although $p=1/5$ is the same as Rossby, the dependence on $\Pr$ in~\eqref{shishMess} differs from that of Rossby in~\eqref{RossbyClassico}.
Moreover Rossby was concerned with $Pr \gg 1$, while scaling~\eqref{shishMess} ostensibly applies with  $\Pr \ll 1$.
Referring to $\mathrm{I}_\ell$ as Rossby scaling  is a misnomer: the phase diagram. of \cite{Shish16}  does not contain a region corresponding to the original $\Pr \gg 1$ Rossby scaling in~\eqref{RossbyClassico}.

\subsection{BL balance of terms in the vorticity equation~\eqref{vorty7}\label{2Dbal}}

All solutions in this work have $\Pr = 1$ and  we have nothing to say regarding the different $\Pr$-exponents in~\eqref{RossbyClassico}, \eqref{GGH29}, and~\eqref{shishMess}.
We can, however, examine the balance of terms in the spanwise-averaged vorticity equation~\eqref{vorty7} and so assess the assumptions made by the three scaling arguments summarized above.
To compactly present these results we further $x$-average~\eqref{vorty7} over the interval $0 < x / h < 1.85$ at fixed $z$.
By stopping the $x$-average at $x/h=1.85$, we stay away from the moving attachment point of the plume.
(Recall that the midpoint of the domain, which is also the position of maximum surface density, is at $x = 2h$.)
The $x$-average of~\eqref{vorty7} also removes small-scale fluctuations that dominate the BL at the highest values of $\Ra$, e.g.~see figure~\ref{Fig5}(c) and (d).
There is no time averaging: this partial $x$-average is applied to a single snapshot. 

In the following discussion we nominally define  the main boundary layer (MBL hereafter) as the region within $2\dafive$ from the top surface:
\beq
  h - 2 \dafive \leq z \leq h \com
  \label{MBLdef}
\eeq
and the inner boundary layer (IBL hereafter) as the region within $2\dafour$ from the top surface:
\beq
  h - 2 \dafour \leq z \leq h \per
  \label{IBLdef}
\eeq

\subsubsection{The 2D solutions}

\begin{figure}
\centering
\includegraphics[width=1.0\textwidth]{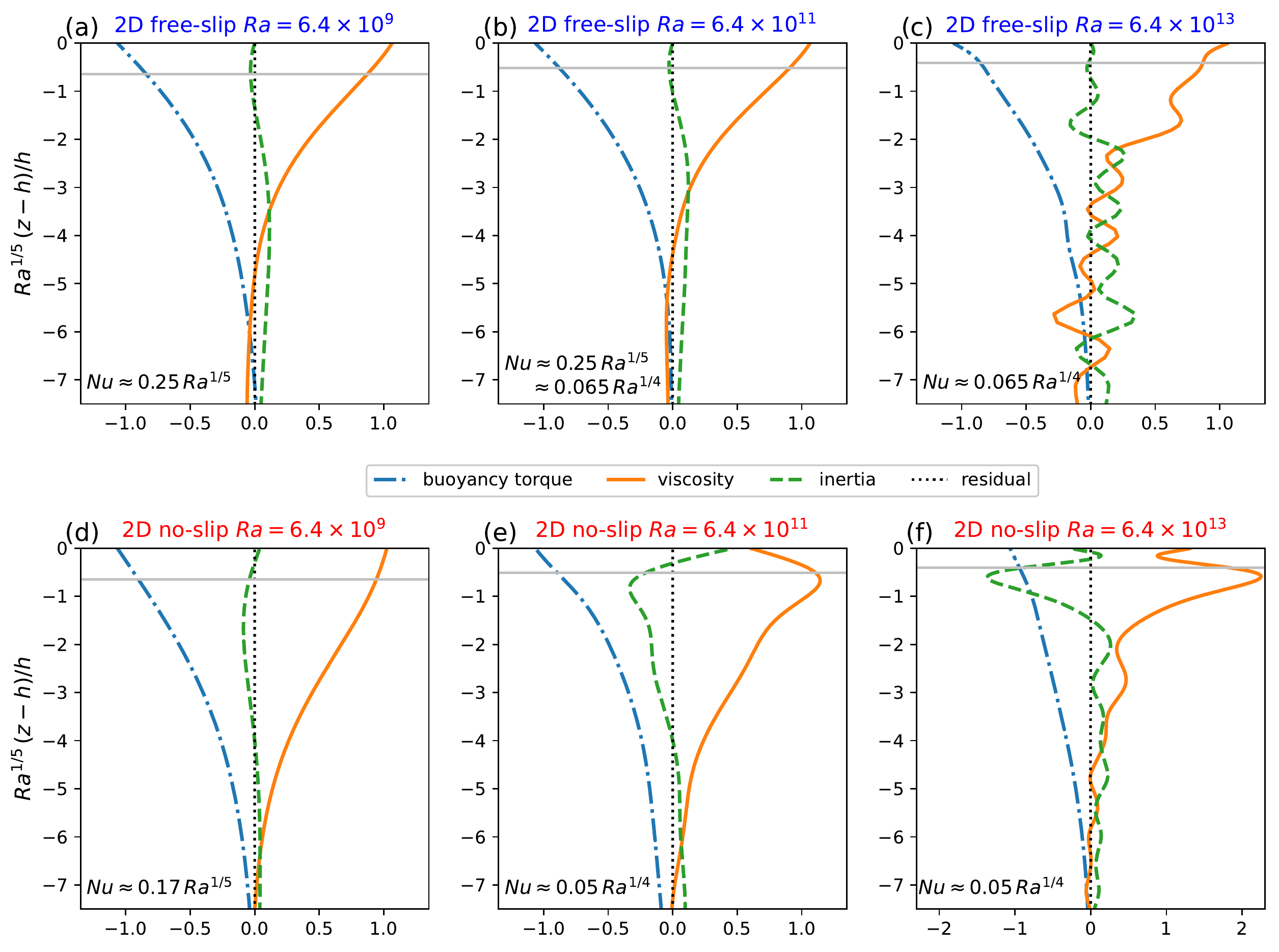}
\caption{Three terms in the 2D version of the  vorticity equation~\eqref{vorty7}. The top row shows \twoDFS{}   solutions and bottom row shows \twoDNS{}  solutions. The abscissa scale is expanded in~(f). Solutions above are in the scaling regimes indicated in the bottom left corner of each panel; the solution in~(b) is at a transition between one-fifth and one-fourth. 
All terms are non-dimensionalized using $\bstar/h$, e.g.~buoyancy torque is $b_x\,h/\bstar$. Horizontal grey lines indicate the distance $2\dafour$ below the top surface $z=h$.}
\label{Fig6}
\end{figure}

Figure~\ref{Fig6} shows the balance of terms in the 2D version of ~\eqref{vorty7}. There is no Reynolds stress torque (RST) in~2D.
 
The MBL in figure~\ref{Fig6}(a) and (b) is exemplary of the two-term Rossby balance between buoyancy torque and viscosity. In fact the  \twoDFS{} solutions in the top row of figure~\ref{Fig6} are characterized very simply: inertia is subdominant in the \twoDFS{} MBL. This even applies to the $\Ra = 6.4 \times 10^{13}$ solution in figure~\ref{Fig6}(c), which is in the $p=1/4$ regime with  $\Kfour = 0.065$ in figure~\ref{Fig4}(b). Below the MBL the situation is more complicated: in figure~\ref{Fig6}(c), there is cancellation between viscosity and inertia. In this region, below the MBL, buoyancy torque is not negligible: there is a three-term balance between buoyancy torque, viscosity and inertial in the \twoDFS{} vorticity equation.
The  \twoDFS{}  solutions are remarkable because Rossby's balance applies throughout the MBL up to the maximum  $\Ra =6.4 \times 10^{13}$ i.e. Rossby's balance applies even in the   $p=1/4$ non-Rossby scaling regime.

For the \twoDNS{} solutions in the bottom row of figure~\ref{Fig6}, the situation is more complicated.
The base of the IBL in~\eqref{IBLdef} is indicated the grey horizontal lines in figure~\ref{Fig6}. It is striking that in figure~\ref{Fig6}(e) and~(f), inertia is largest within the IBL and that there is a three-term balance in the \twoDNS{}  vorticity equation.

Examination of the vorticity equation~\eqref{vorty7} evaluated at the surface $z = h$ provides a rationalization for the different role of inertia in \twoDFS{} versus \twoDNS.
At $z=h$ the inertia term in~\eqref{vorty7} is
\begin{align}
  \text{inertia}&\defn  \zeta_t + \psi_x \zeta_z  - \psi_z \zeta_x \com \\
  & = \begin{cases} \, 0 \com \qquad \text{with \FS{} boundary conditions ($\zeta =0$);}\\
  \zeta_t \com \qquad \text{with \NS{} boundary conditions ($\psi_x = \psi_z = 0$).}
  \end{cases}
  \label{clearcut}
\end{align}
In the \twoDFS{} case inertia is identically zero at the surface and  there must therefore be a balance between the two remaining terms in the 2D version of~\eqref{vorty7}, i.e. at $z = h$ viscosity must balance buoyancy torque.
In fact this Rossby balance applies throughout the \twoDFS{} MBL. 

In the \twoDNS{} case, inertia in~\eqref{vorty7} might be  non-zero if there are temporal fluctuations so that $\zeta_t \neq 0$. In figure~\ref{Fig6}(d) these $\zeta_t$ fluctuations are small, and so is inertia. As the $\Ra$ is increased, e.g.~ in figure~\ref{Fig6}(e) and~(f), the temporal fluctuations are larger and inertia at the surface $z=h$ becomes comparable to buoyancy torque and viscosity.

The argument above requires qualification in the special -- but popular -- case of piecewise constant $\bs(x)$, e.g.~$\bs(x) = \pm \bstar$, with a discontinuous jump at the middle of the domain.
While~\eqref{clearcut} is valid, the buoyancy torque $\hat b_x$ at the surface vanishes, except at the discontinuity.
In this case the 2D version of~\eqref{vorty7} everywhere except the discontinuity, when evaluated at the wall, collapses to $\zeta_t = \nu \lap \zeta$ and no conclusion can be drawn.
This observation is perhaps an indication that piecewise constant buoyancy profiles might be qualitatively different from smoothly varying buoyancy profiles.


\subsubsection{The 3D solutions}

\begin{figure}
\centering
\includegraphics[width=1.0\textwidth]{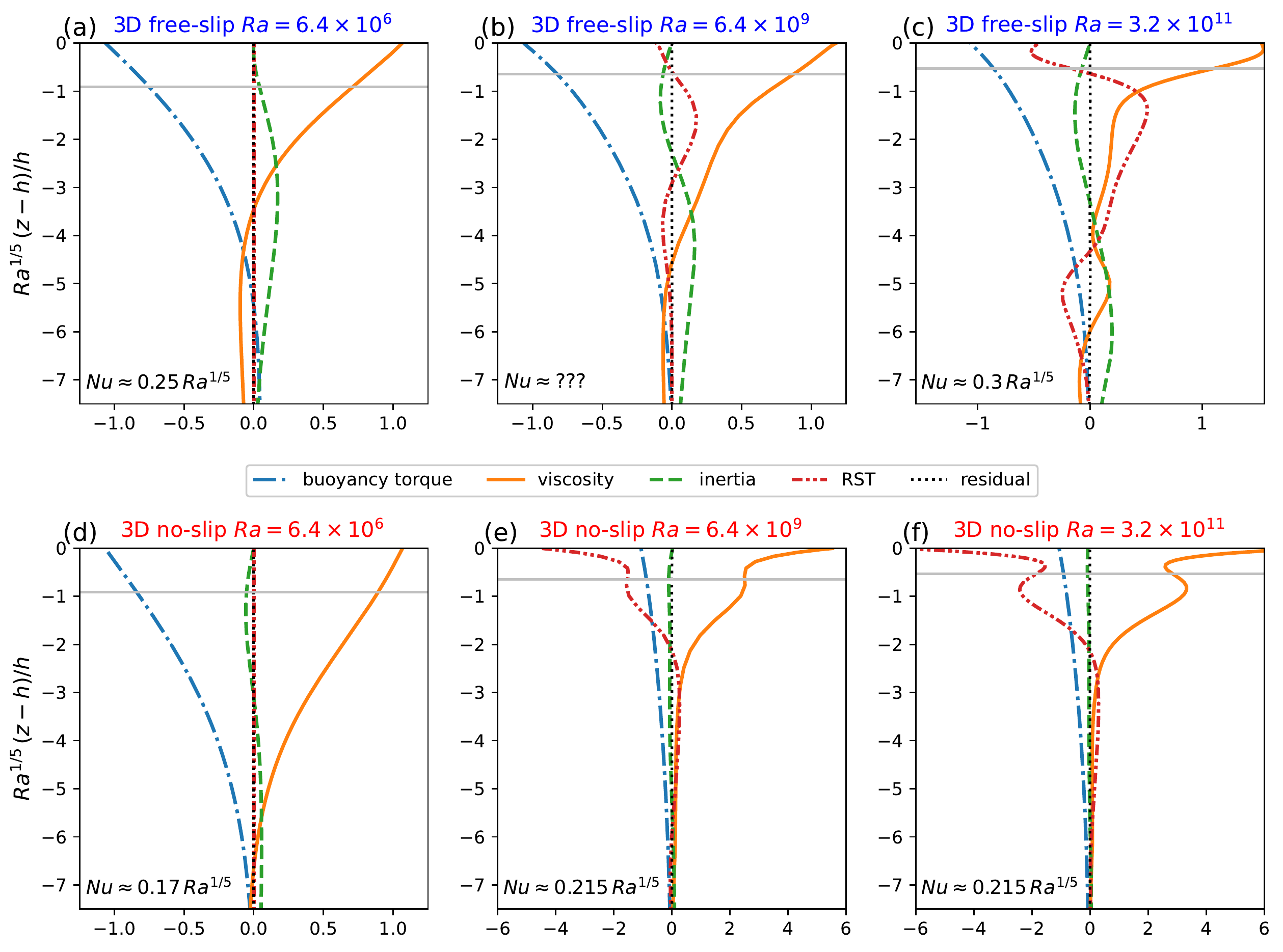}
\caption{Four terms in the vorticity equation~\eqref{vorty7}, averaged over $0 \le x / h \le 1.85$, and calculated from snapshots of the 3D solutions. The top row shows \threeDFS{} solutions and bottom row shows  \threeDNS{} solutions. The abscissa scale varies between the panels. Solutions  above are in the scaling regimes indicated in the bottom left corner of each panel. In panel (b) $Nu =???$ indicates that this solution in the transition between the two $p=1/5$ regimes in (a) and (c). All terms are non-dimensionalized using $\bstar / h$, e.g.~buoyancy torque is $b_x \, h / \bstar$.  Horizontal grey lines indicate the distance $2\dafour$ below the surface.}
\label{Fig7}
\end{figure}


In figure~\ref{Fig7} we turn to the 3D solutions and analyze the balance of terms in the spanwise-averaged vorticity equation~\eqref{vorty7}. In 3D the  term RST in~\eqref{vorty7} might be non-zero.

We use the partial $x$-average described in the previous section. From the outset we draw attention to a simple result: within the MBL of figure~\ref{Fig7}, inertia in the 3D vorticity equation~\eqref{vorty7} is always subdominant relative to the other three terms. The unimportance of MBL inertia is striking in the \NS{} solutions in bottom row of figure~\ref{Fig7} -- but it is also a not-bad approximation for the \FS{} solutions in the top row.

The \threeDFS{} solution in figure~\ref{Fig7}(a) is at the low Rayleigh number $6.4\times 10^6$ and is effectively 2D and steady. This solution, which is in the middle of a four-decade $p=1/5$ regime with $\Kfive = 0.25$, exhibits a two-term Rossby balance between viscosity and buoyancy torque. Inertia becomes as large as viscosity just below the MBL.

Figure~\ref{Fig7}(d) shows a \NS{}  solution from the middle of the first $p=1/5$ scaling regime -- with $\Kfive =0.17$ in figure~\ref{Fig4}(a). Figures~\ref{Fig7}(e,f) show  solutions from the second $p=1/5$ scaling regime with $\Kfive  = 0.215$. In figure~\ref{Fig7}(d) there is  a Rossby  balance between viscosity and buoyancy torque, while in figures~\ref{Fig7}(e,f) the situation is complicated: close to the top surface, within the IBL,  there is a two-term balance between RST and viscosity. Further from the surface, but within the MBL, there is a three-term balance involving viscosity, buoyancy torque and RST.
 
The  \threeDNS{} solutions in the bottom row of figures~\ref{Fig7} confirm the result of \cite{G14}: the difference between the two \threeDNS{}  $p=1/5$ regimes results from a transition between a low-$\Ra$ Rossby balance in panel~(a) to an unsteady 3D BL flow involving large RST in panels~(e) and~(f).  


\subsubsection{Summary and conclusion based on the BL balance of terms}

The twelve panels in figures~\ref{Fig6} and~\ref{Fig7} summarize results from twelve solutions, with both \FS{}  and \NS{}  boundary conditions, and both 2D and 3D configurations.
The main conclusions are:
\begin{enumerate}
\item[\textit{(i)}] the structure of the buoyancy torque is the same in the twelve cases;
\item[\textit{(ii)}]  in the $p=1/5$ and $p=1/4$ regimes  the thickness of the buoyancy-torque BL is $\dafive$;
\item[\textit{(iii)}] buoyancy torque is of leading importance in ten of the twelve cases, and in seven cases Rossby's two-term balance between buoyancy torque and viscosity holds throughout the MBL;
\item[\textit{(iv)}] the buoyancy torque is subdominant only for the \threeDNS{}  solutions at the highest $\Ra$'s, resulting in  a two-term balance between viscosity and RST;
\item[\textit{(v)}] in the high-$\Ra$ \NS{} solutions, there is  a thin $\dafour$-IBL nested within the   thicker $\dafive$-MBL.
\end{enumerate}

Points \textit{(i)} and \textit{(ii)}  indicate that main  features of  the buoyancy field are insensitive to the balance of terms in the vorticity equation, and to the viscous boundary condition and to dimensionality. Point \textit{(iii)} indicates that Rossby's $\Pr \gg 1$ balance of terms can apply at moderately large $\Ra$ even with $\Pr=1$.

 All solutions exhibit the same $\Nu$--$\Ra$ scaling phenomenology, even though the balance of terms in the vorticity equation changes with viscous boundary condition and dimensionality. For example, in the 2D results in figure~\ref{Fig6}, inertia plays a different role in the top row (\FS) than in the bottom row (\NS). Yet the  transition from $p=1/5$  to~$p=1/4$ is qualitatively the same for the two viscous boundary conditions.


\section{Nested boundary layers \label{oneQuarter}}

\citet{tsaiSheard20} investigate \twoDNS{} HC with $\Pr=6.14$.
With $\Ra > 10^{10}$, and provided that the imposed surface buoyancy varies linearly with the horizontal coordinate $x$, \citet{tsaiSheard20} report the $p=1/4$ scaling~\eqref{oneFourth} extending over four decades of $\Ra$. Here we also find $p=1/4$ in the \twoDFS{} and  \twoDNS{} solution suites. \citet{tsaiSheard20} speculate that the $p=1/4$ scaling  might correspond to a regime proposed by \cite{Shish16} in their phase diagram of the $(\Ra, \Pr)$ parameter plane.
In this scheme the $(\Ra,\Pr)$-plane is partitioned into seven regions and the exponent $1/4$ is located in regions denoted  III$_{\infty}$, IV$_u$, and I$^*_\ell$.
But we now show that defining features of III$_{\infty}$, IV$_u$, and I$^*_\ell$ do not agree with the numerical solutions.
We conclude that $p=1/4$ found here, and likely also in \citet{tsaiSheard20}, is not in agreement with any region of the phase diagram of \citet{Shish16}.

\subsection{Partitioning of buoyancy dissipation $\chi$ between BL and interior}
 
\begin{figure}
  \centering
  \includegraphics[width=0.8\textwidth]{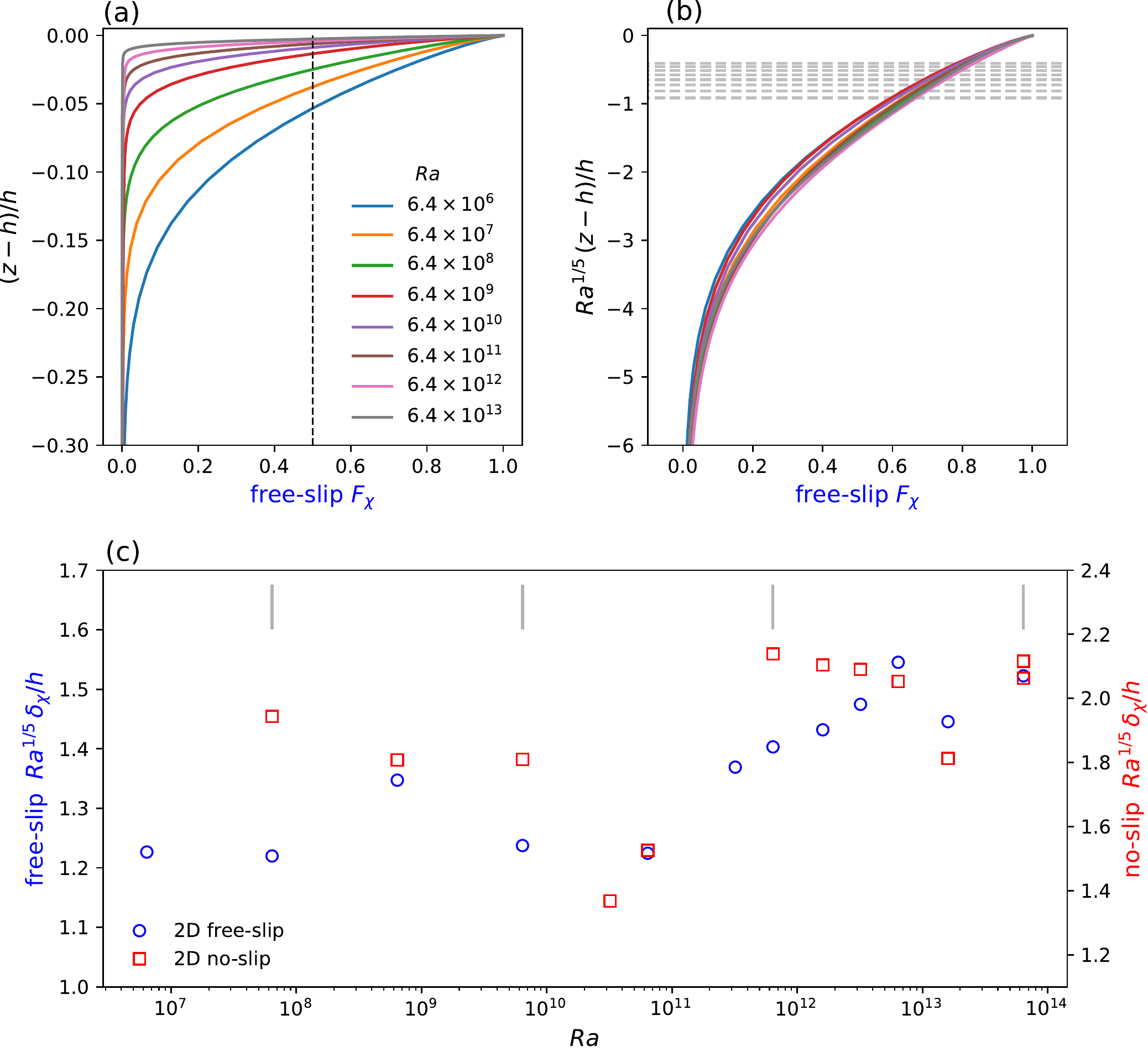}
  \caption{(a) The function $F_{\chi}(z)$ defined in~\eqref{Fdefz} for the \twoDFS{} solutions.  A BL thickness, $\delta_\chi$, is defined as the distance from the top at which $F_\chi(z) = 1/2$ (dashed vertical lines in panel~(a)). (b)~Same as~(a) but with the vertical axis rescaled with $\Ra^{1/5}$. Horizontal grey dashed lines indicate the distance $2 \dafour$ below the top surface $z = h$. Panels~(c) and~(d) are the same as~(a) and~(b) respectively but for the \twoDNS{} solutions. (e)~The compensated BL thickness, $\Ra^{1/5} \delta_\chi / h$, as a function of $\Ra$ for both the \twoDFS{} and the \twoDNS{} solutions. The four vertical grey line segments mark the $\Ra$'s of the solutions in figures~\ref{Fig1} and~\ref{Fig2}.}
  \label{Fig8}
\end{figure}

A main characteristic distinguishing the various regimes by \cite{Shish16} is the partitioning of kinetic energy dissipation, $\varepsilon$, and buoyancy variance dissipation, $\chi$, between the BL and the interior of the domain.
To quantify the partitioning of $\chi$ we introduce the function
\beq
  F_{\chi}(z) \defn {\frac{\kappa}{h\chi} \int_0^z \overline{|\grad b|^2} \, \dd z'} \com
  \label{Fdefz}
\eeq
where the overbar denotes an $(x, y, t)$-average.
$F_\chi(z)$ in figure~\ref{Fig8} increases monotonically from $0$ to $1$ with $z/h$ and indicates the fraction of buoyancy-variance dissipation below the level $z$.
Figures~\ref{Fig8}(a) and (c) show that~$\chi$ is increasingly localized within a BL as $\Ra \to \infty$.

A main characteristic of regions III$_{\infty}$ and IV$_u$ in the phase diagram of \cite{Shish16} is that $\chi$ is dominant in the interior of the domain.
Thus figures~\ref{Fig8}(a) and (c) disqualify regions~III$_{\infty}$ and~IV$_u$.
The remaining possibility with exponent $1/4$ is the $\Pr \gg 1$ region~$\mathrm{I}_\ell^*$, characterized by a momentum BL that is much thicker than the buoyancy diffusion BL.
But region $\mathrm{I}_\ell^*$ is located at moderate values of $\Ra$ in the phase diagram so that $1/4$ is the first exponent encountered if $\Pr$ is fixed and $\Ra$ is increased from small values.
In our 2D solutions, however, we first find  $p=1/5$ scaling~\eqref{RossbyScaling}, which is replaced at higher $\Ra$ by $p=1/4$ in~\eqref{oneFourth}.
This is also the case in the study of \citet{tsaiSheard20}: first $1/5$ and then, at higher $\Ra$, $1/4$.
We conclude that  $p=1/4$ in the 2D solution suites is not related to regions III$_{\infty}$, IV$_u$ and I$^*_\ell$ of the phase diagram of \cite{Shish16}.
In section~\ref{2DT} we seek an alternative explanation for the $p=1/4$ regimes in figure~\ref{Fig4}(b).

To extract more information from $F_{\chi}(z)$, we scale the $z$-axis with $\dafive$ and re-plot the results from figures~\ref{Fig8}(b) and (d).
The curves now fall largely on top of each other, indicating that the function $F_{\chi}(z)$ expresses the BL thickness $\dafive$, even if $\Nu \sim \Ra^{1/4}$.
To quantify this, we define a BL thickness, $\delta_\chi$, by determining the level at which $F_\chi(z) = 1/2$: see the dashed vertical line in figures~\ref{Fig8}(a) and (c).
The compensated plot in figure~\ref{Fig8}(e) then shows that
\beq
  \delta_\chi \approx K_\chi \dafive \per
  \label{myst}
\eeq
The constant $K_\chi$ in~\eqref{myst} is about $1.5$ for the \FS{} solutions and $1.8$ for the \NS{} solutions.

The scaling in~\eqref{myst} applies in \textit{both} the $p=1/5$ regime~\eqref{RossbyScaling} and the $p=1/4$ regime~\eqref{oneFourth}.
Yet the reasoning in section~\ref{RossbySec}, leading to 
\beq
  \Nu \sim \frac{\h}{\dab} \com
  \label{NulAgain}
\eeq
underpins all scaling arguments and seems inescapable.
It must be that in the $p=1/4$ regime~\eqref{oneFourth}, the buoyancy BL has a double-layer structure: there is a thin BL, with thickness $\dab = \dafour$, embedded within the thicker $\delta_{\chi}$-BL in~\eqref{myst}.
In the next section we show the existence of such a double BL.

(The scatter of $K_\chi$ in figure~\ref{Fig8}(c) might be considered uncomfortably large.
Note, however, that $\Ra$ is varied by seven decades.
This large range encompasses the transition from steady to strongly time-dependent 2D flows.
At $\Ra = 6.4 \times 10^{13}$, $\Ra^{-1/4}$ is smaller than $\Ra^{-1/5}$ by a factor of 5, which is much greater than the $\pm 20\%$ scatter in figure~\ref{Fig8}(c).)

\subsection{The $\dafour$-boundary layer}

To identify the $\dafour$-BL in our solutions, and show consistency with~\eqref{NulAgain} in the $p=1/4$ regime, we notice that with the sinusoidal surface buoyancy in~\eqref{sbuoy8}, the surface Nusselt number~\eqref{surfNussl} is
\begin{align}
  \Nus = b'_{1}(h) \Big \slash b'_{\text{diff} 1}(h) \com
  \label{sBL3}
\end{align}
where above the prime denotes a $z$-derivative and 
\beq
  b_1(z) \defn 2 \, \overline{\cos k x \, b(x,y,z,t)} \per
  \label{eq:b_1dfn}
\eeq
Using~\eqref{sBL3} we obtain an alternative expression for $\dab$:
\beq
  \dab \defn \bstar / b'_1(h) \com
  \label{deltasdef}
\eeq
where $\bstar$ is the amplitude of the sinusoidal surface buoyancy in~\eqref{sbuoy8}.
The numerator in~\eqref{deltasdef} is appropriate because  $\bstar = b_1(h) = b_{\text{diff} 1}(h)$.

\begin{figure}
  \centering
  \includegraphics[width=1.0\textwidth]{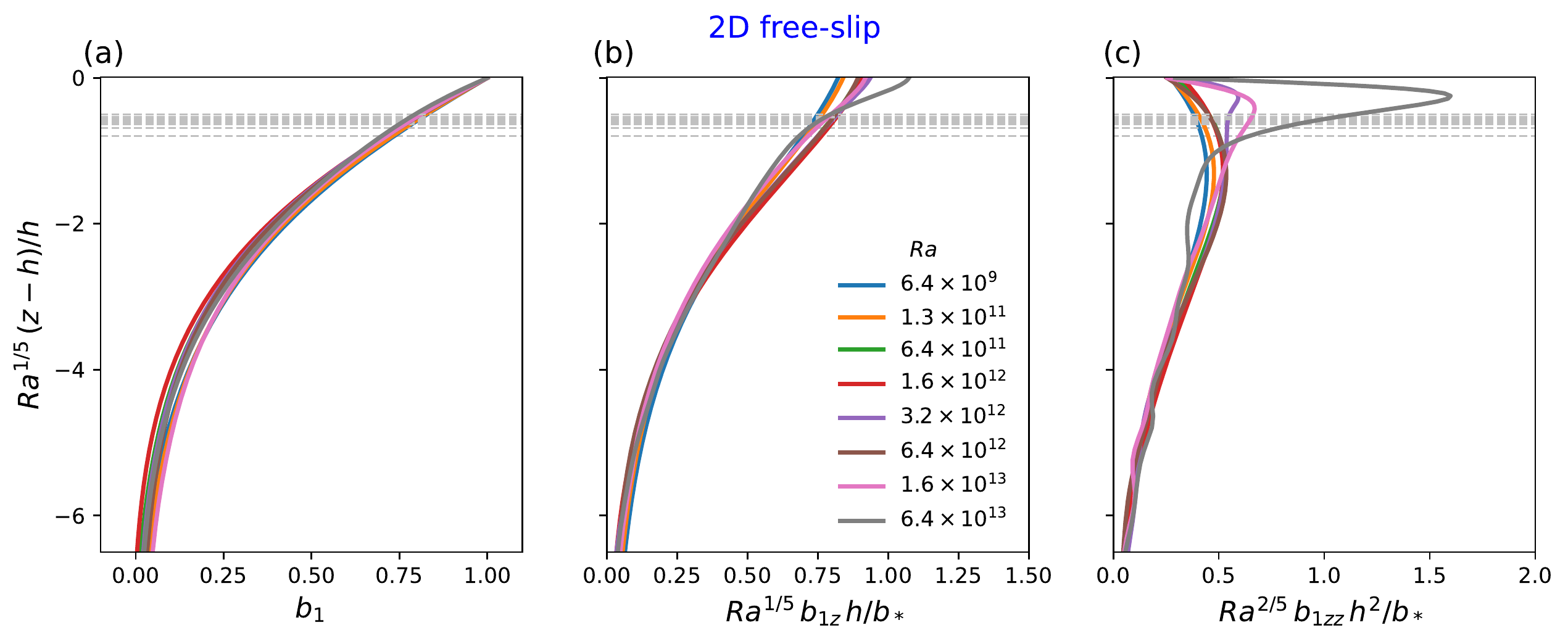}
  \caption{The structures of (a)~$b_1(z)$, (b)~$b'_1(z)$, and (c)~$b''_1(z)$ for \twoDFS{} solutions at various~$\Ra$. $b_1(z)$ was obtained here  from the final snapshot  without any time-averaging. Horizontal grey dashed lines indicate the distance $2 \dafour$ below the top surface $z=h$.}
  \label{Fig9}
\end{figure}

Figure~\ref{Fig9} shows $b_1(z)$, and the first two derivatives of $b_1(z)$
The overline in~\eqref{eq:b_1dfn} indicates both a horizontal and temporal average.
Figure~\ref{Fig9} is based on the horizontal average of single snapshots of the buoyancy field at the final time.
The inner BL, with thickness $\dafour$, is not visible in $b_1(z)$ in figure~\ref{Fig9}(a).
But the higher derivatives of $b_1(z)$ in the panels (b) and~(c) reveal the scale $\dafour$.
In particular, the $p=1/4$ scaling~\eqref{oneFourth} results from the increase in $b'_1(h)$, evident in figure~\ref{Fig9}(b) as $\Ra$ increases.
The maximum of $b_1''(z)$ in figure~\ref{Fig9}(c) appears only in the $p=1/4$ regime~\eqref{oneFourth}.

\begin{figure}
  \centering
  \includegraphics[width=0.9\textwidth]{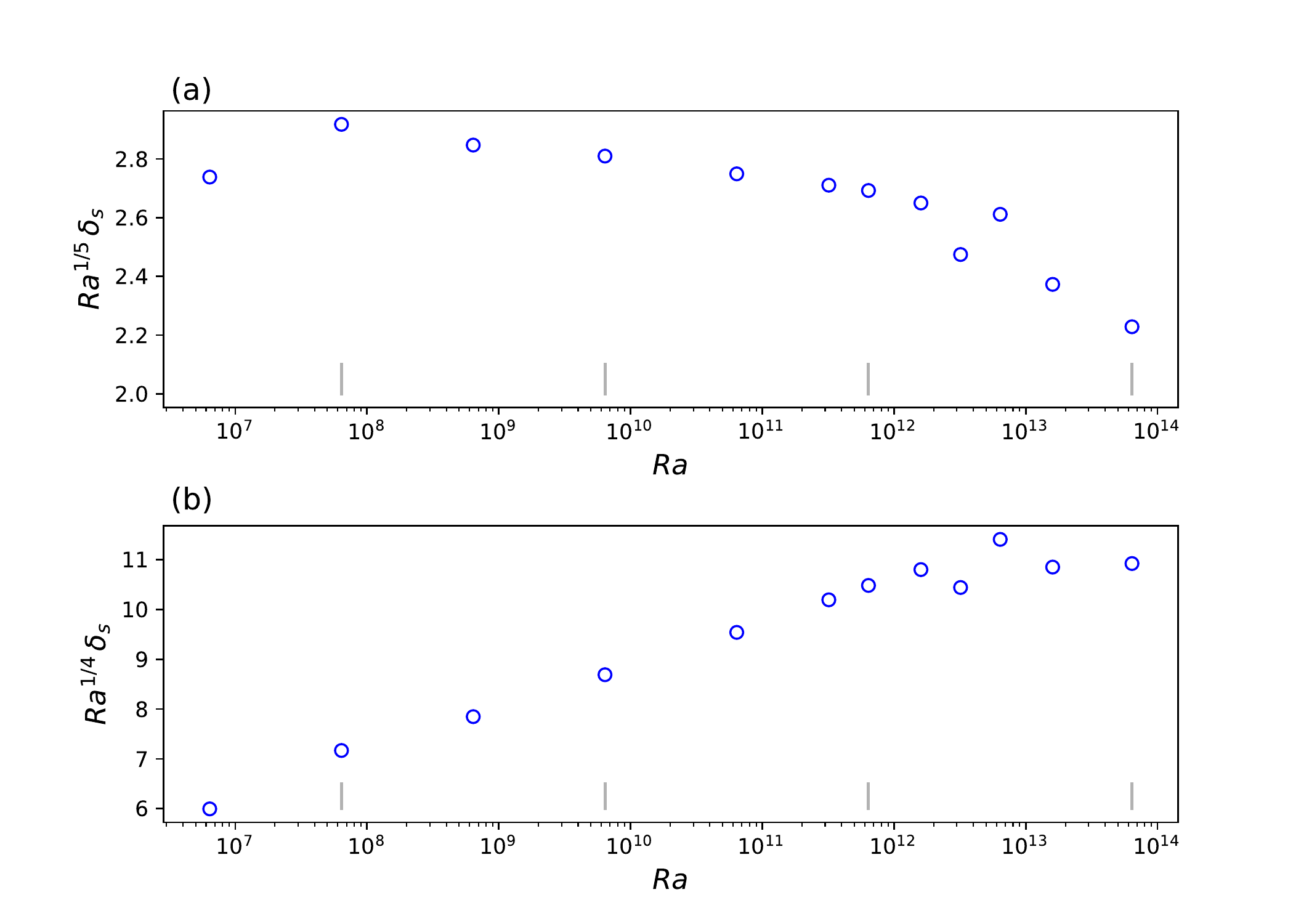}
  \caption{The compensated surface BL thickness, $\dab$ in~\eqref{deltasdef}, for the \twoDFS{} suite. In panel~(a) $\dab$ is compensated by $\Ra^{-1/5}$ and in~(b) by $\Ra^{-1/4}$. The transition between the $p=1/5$ scaling~\eqref{RossbyScaling} and the $p=1/4$ scaling~\eqref{oneFourth} is at about $\Ra = 10^{12}$. Vertical grey line segments mark the $\Ra$ values corresponding to the solutions in figures~\ref{Fig1} and~\ref{Fig2}.}
  \label{Fig10}
\end{figure}

Figure~\ref{Fig10}(a) shows $\dab$, diagnosed from~\eqref{deltasdef}, and compensated by $\Ra^{1/5}$. For $\Ra$'s corresponding to $p=1/5$ scaling  this compensated  $\dab$ in the narrow range $2.6$ and~$2.9$. Thus with $p=1/5$,  both $\dab$ and $\delta_\chi$ vary as  $\dafive$.
Figure~\ref{Fig10}(b) shows that at the four or five highest values of $\Ra$, corresponding to the $p=1/4$ scaling, $\dab$ compensated with $\Ra^{1/4}$ varies between about $10.5$ and $11.4$. We conclude that in the $p=1/4$ regime  $\dab \sim \dafour$, but $\delta_\chi \sim \dafive$. 


\section{A scaling argument for $\Nu \sim \Ra^{1/4}$ \label{2DT}}

In this section we present a scaling argument applicable to the $p=1/4$ regime of horizontal convection.
Although our numerical solutions revealed the $p=1/4$ scaling  only in the 2D cases, we still hold hope that $p=1/4$ might also emerge in 3D at sufficiently high $\Ra$.
With the 3D case in mind, we propose an overarching explanation for the $p=1/4$ scaling -- independent of boundary conditions, dimensionality and   2D vortex-gas phenomenology.

The $p=1/4$ regime requires an inner buoyancy BL with thickness $\dab \sim \dafour$.
In discussing this inner BL it is helpful to keep figure~\ref{Fig5}(d) in mind: the $\dafour$-BL is identified by the dashed grey line.
Think of this inner BL as a laminar sub-layer, stirred by the outer flow in the much thicker $\dafive$-BL.
The overarching explanation alluded to above is that the thickness of the laminar sub-layer is related to the Kolmogorov and Batchelor length scales
\beq
  \etaK = \left( \frac{\nu^3 }{ \varepsilon} \right)^{1/4} \andd \etaB = \left( \frac{\kappa^2 \nu }{ \varepsilon} \right)^{1/4} \per
  \label{KB3}
\eeq
These length scales are identified as the smallest scales of fluctuations in momentum and buoyancy that can survive before the damping by viscosity $\nu$ and diffusion $\kappa$ is overwhelming.
By analogy, the HC laminar sub-layer with thickness $\dab$ is the thinnest BL that can survive in a horizontal-convective  flow that is supplied with kinetic energy at a rate $\varepsilon$.

 (In the arguments of Kolmogorov and Batchelor the viscous dissipation rate  $\varepsilon$ is also the energy cascade rate in a 3D inertial range.
 This interpretation of $\varepsilon$ cannot apply to 2D HC: there is no vortex stretching in a 2D flow and therefore no forward cascade of energy.
 We argue instead that the laminar sub-layer thickness is determined by $\varepsilon$ as the most basic measure of forcing strength and by the molecular parameters $\nu$ and $\kappa$.
 Thus $\dab \sim (\nu^p\kappa^q/\varepsilon)^{1/4}$, with $p + q = 3$, is dimensionally acceptable; $\etaK$ and $\etaB$ are the most prominent members of this family.
 Saying more would would require varying $\Pr$ which is beyond our scope here.)

Following the  scaling arguments reviewed section~\ref{RossbySec}, we assume that $\bar b(0) \approx - \bstar$.
Then, once again, the energy power integral~\eqref{PY2.5} implies that
\beq
  \varepsilon \sim \frac{\kappa \bstar}{h} \per
  \label{PY101}
\eeq 
With $\varepsilon$ in~\eqref{PY101}, $\etaK$ and $\etaB$ in~\eqref{KB3} can be written as
\beq
  \frac{\etaK}{h} = \Pr^{1/2} \Ra^{-1/4} \andd \frac{\etaB}{h} = \Pr^0 \Ra^{-1/4} \per
  \label{etaKB3}
\eeq 
Thus if the laminar sub-layer has thickness $\dab \sim \etaK$, or perhaps $\dab \sim \etaB$, then from~\eqref{Nul} $\Nu \sim \Ra^{1/4}$.

\section{Dimensionality and mean square vorticity \label{3Dffs}}

With periodicity in the spanwise  direction, 2D solutions  are also 3D solutions. At sufficiently high $\Ra$, and with sufficiently large $\elly$, these 2D solutions are unstable to perturbations in the spanwise dimension \citep{G14,Tsai16,PSW17}. Apart from the possibility of non-zero RST in~\eqref{vorty7} the scaling arguments summarized in section~\ref{RossbySec} make no reference  to the development of 3D flow.
To appreciate the limitations of scaling  arguments it is interesting to compare 2D and 3D solutions at the same $\Ra$. Comparison of 3D with 2D  is also relevant to the nature of turbulence in HC \citep{PY02, SW, Shish16}, e.g., turbulent amplification of vorticity by vortex stretching does not operate in 2D. 

Once 3D motion is activated by instabilities one expects initiation of vortex stretching and the development of other associated qualitative differences between 2D and 3D HC.
But this cannot happen in HC: in this section we show using the mechanical energy power integral~\eqref{PY2.5} that in the limit $\Ra \to \infty$ vorticity amplification does not occur in the 3D HC.
To motivate this proof we first note the rather small quantitative differences between 2D and 3D HC at~$\Ra = 3.2 \times 10^{11}$.

\begin{figure}
  \centering
  \includegraphics[width = 0.9 \textwidth]{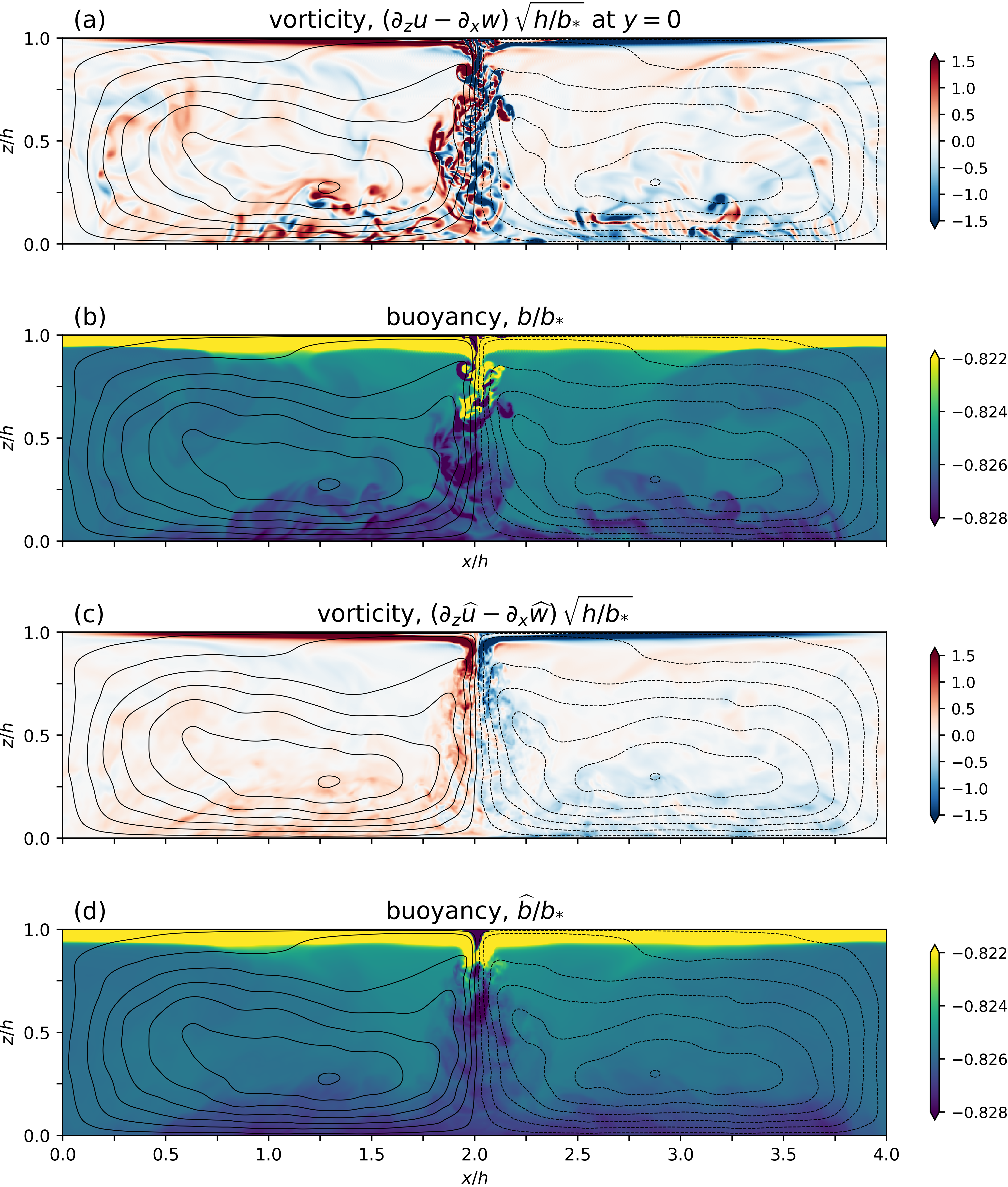}
  \caption{Panels (a) and~(b) show a $y$-slice of snapshots of the \threeDFS{} solution at $\Ra = 3.20 \times 10^{11}$. The spanwise averages of the snapshots of vorticity and buoyancy are shown in panels~(c) and~(d). In all panels, the contours are streamlines computed from a spanwise averaged snapshot at the final time. At the top surface $-1 \leq b / \bstar \leq +1$; the narrow range of the buoyancy color scale makes the small interior variations visible.}
  \label{Fig11}
\end{figure}

In table~\ref{Table1} it is remarkable that the  3D solutions have less than a 20\% enhancement of  $\Nu$ above that of the 2D solutions at the same $\Ra$.
It  is difficult to believe that this  modest 20\% enhancement extends to geophysical $\Ra$'s, which exceed the largest 3D $\Ra$'s in table~\ref{Table1} by perhaps $10^{12}$.
If the 3D increase in $\Nu$ remains at 20\%, then relatively inexpensive 2D numerical solutions would provide useful estimates of 3D heat transport.

(It is likely that the 2D $\Nu$ is always less than the 3D $\Nu$ i.e. the onset of 3D motion cannot be  accompanied by a decrease in  heat transport.
A proof of this conjecture would be useful.) 

To further emphasize the small 3D increase in $\Nu$, notice that boundary conditions have a larger quantitative effect on heat transport than dimensionality: in figure~\ref{Fig4} the \twoDFS{} solutions have larger $\Nu$ than the \threeDNS{} solutions.

Table~\ref{billzTab} summarizes gross measures of the departures from the 2D spanwise-averaged circulation defined in~\eqref{span3}.
For both both our highest-$\Ra$ \threeDNS{} and \threeDFS{} solutions, about two-thirds of the kinetic energy is in the spanwise-averaged flow.
In the third column the component of buoyancy gradient in the spanwise direction ($b_y$) contributes only less than $2$\% to the buoyancy dissipation $\chi$.
The only statistic that is dominated by departures from the spanwise average is mechanical energy dissipation, $\varepsilon$, in the fourth column.
Thus table~\ref{billzTab} -- particularly the third column -- supports the view that the 3D $Nu$  at $\Ra = 3.20 \times 10^{11}$ is largely determined by the 2D spanwise averaged circulation.

\begin{figure}
  \centering
  \includegraphics[width = 0.9 \textwidth]{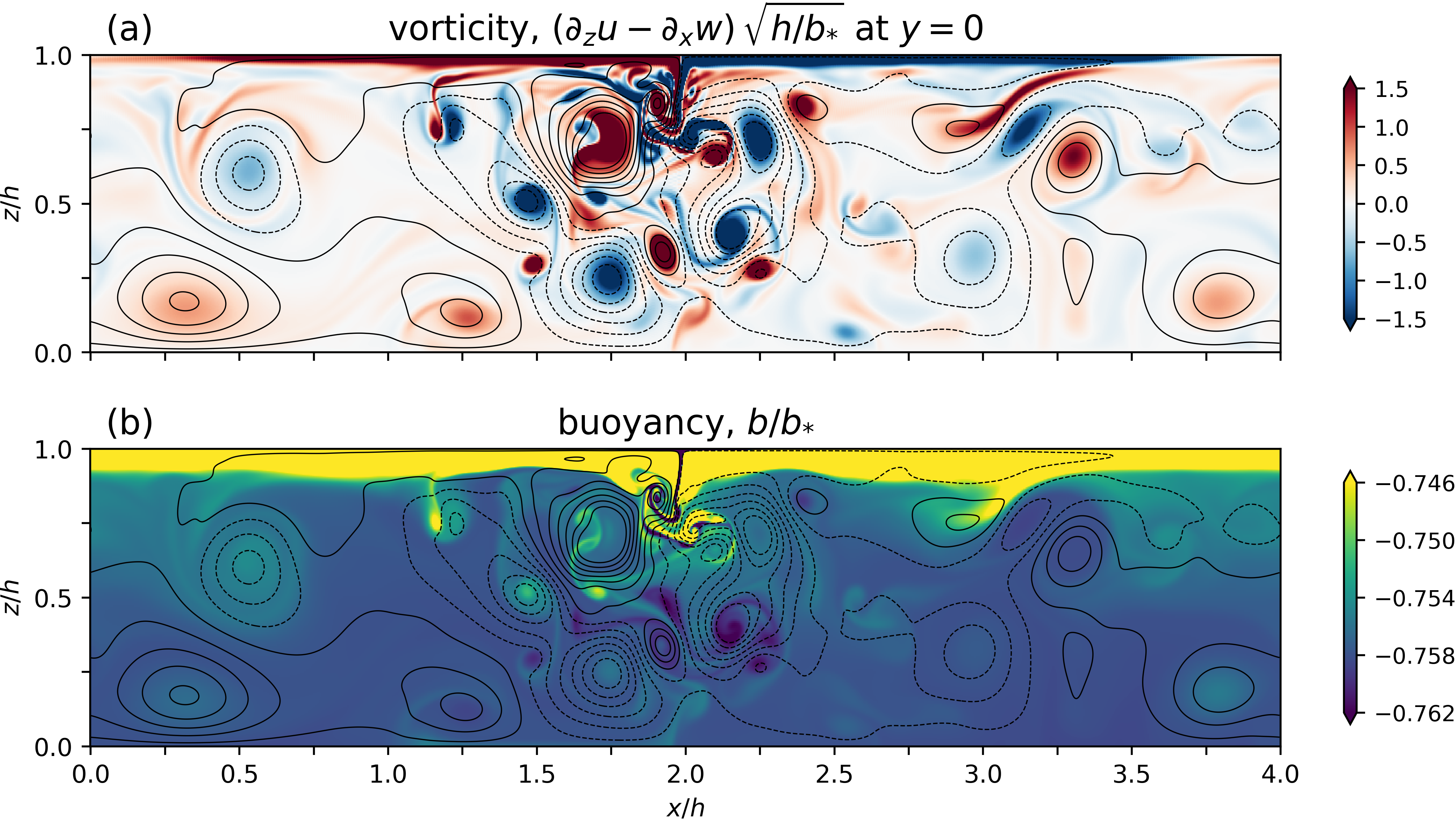}
  \caption{Snapshot of the \twoDFS{} solution at $\Ra = 3.20 \times 10^{11}$. The contours are streamlines. The narrow range of the buoyancy color scale makes the small interior variations visible.}
   \label{Fig12}
\end{figure}

Figure~\ref{Fig11}(a) shows that there is no inertial cascade in the interior of the \threeDFS{}  solution. An inertial cascade is characterized by a kinetic energy spectrum $\sim \varepsilon^{2/3} k^{-5/3}$, or a vorticity spectrum $\sim \varepsilon^{2/3} k^{+1/3}$.
The ultra-violet vorticity divergence is cut-off at a wavenumber of order $\etaK^{-1}$, where $\etaK\sim \dafour$ is the  Kolmogorov length.
But  the snapshot in figure~\ref{Fig11}(a) shows that the interior vorticity is concentrated on length scales very much larger than $\etaK$.
Instead of looking for signatures of 3D turbulence, it is informative to compare figure~\ref{Fig11} with the corresponding \twoDFS{} run in figure~\ref{Fig12}.  

The 2D vortices in figure~\ref{Fig12}(a) have larger length scale and are more intense than the 3D vorticity fluctuations in figures~\ref{Fig11}(a) and (c).
The larger scale of the 2D vortices might result from vortex mergers occurring during  the disintegration central plume.
These mergers preserve vorticity extrema  \citep{Carnevale1991} so that strong vorticity from the BL -- evident in  figures~\ref{Fig5} (c) and (d) --  is carried into the interior.

The comparison of figure~\ref{Fig11} with~\ref{Fig12} indicates that there is no significant amplification of vorticity in the interior of the 3D solution.
Indeed   the interior vorticity is stronger in the 2D solution than in the 3D solution. Perhaps $\Ra = 3.2\times 10^{11}$ is far too small  for turbulent activation of vortex stretching in the interior of HC? But we now show that  significant 3D vorticity amplification cannot occur in HC even as $\Ra \to \infty$. In other words,  $\la |\bomega|^2 \ra_{3D}$ is almost equal to $\la |\bomega|^2 \ra_{2D}$ even as $\Ra \to \infty$.

\renewcommand{\arraystretch}{1.35}
\begin{table}
\begin{center}
\begin{tabular}{l  c  c c c}
{}  & ${\la |\hat \bu|^2\ra }\slash{\la |\bu|^2\ra}$ \qquad   &
\quad ${\la v^2\ra }\slash{\la |\bu|^2\ra}$ \quad & 
\quad${\kappa \la b_y^2\ra }\slash \chi$\quad & 
\quad $\nu\la \zeta^2\ra\slash \varepsilon $ \quad \\ 
\threeDFS{} & 0.662 & 0.112 	& 0.013 & 0.187 \\
\threeDNS{} & 0.690 & 0.092	& 0.019 & 0.342 \\
\end{tabular}
\end{center}
\caption{Statistics for the 3D solutions at $\Ra = 3.20 \times 10^{11}$. The ratios above were computed from a single snapshot at the final time, i.e., without the benefit of time averaging. These ratios decrease monotonically to zero as $\Ra$ is lowered to the critical value for the onset of 3D motion. In the first column, $\hat \bu =(-\psi_z, 0, \psi_x)$ is the spanwise averaged velocity in~\eqref{overturning} and in the final column $\zeta = - \lap \psi$ is the vorticity of the spanwise-averaged flow.}
\label{billzTab}
\end{table}


With a well known identity, the kinetic energy dissipation can be written as $\varepsilon = \nu \la |\bomega|^2 \ra$ where $\bomega = \curl \bu$ is the vorticity.
Then the mechanical energy power integral~\eqref{PY2.5} rewritten as
\begin{align}
  \frac{\la |\bomega|^2 \ra}{\bstar/h} & = -\frac{\kappa}{\nu} \frac{\bar b(0)}{\bstar} \com\label{PY111}
\end{align}
The left hand side of \eqref{PY111} is a nondimensional mean square vorticity.
The result \eqref{PY111} is independent of $\elly / h$ and can be used to compare a 3D flow with a numerically accessible 2D ``comparison flow'' at the same $\Ra$.
This comparison produces the inequality
\beq
  \frac{\la |\bomega|^2 \ra_{2D}}{\bstar/h}  \leq \frac{\la |\bomega|^2 \ra_{3D}}{\bstar/h}  \leq \frac{\kappa}{\nu}\per
  \label{PY112}
\eeq
The first $\leq $ in \eqref{PY112} is because  if to the contrary $ \la |\bomega|^2 \ra_{2D} > \la |\bomega|^2 \ra_{3D}$ then there is no 3D vorticity amplification and  the proof is finished. The second $\leq $ in \eqref{PY112}  follows from the buoyancy extremum principle $\bar b(0) \geq -\bstar$.


The 2D high $\Ra$ solutions found here already come close to  meeting  the buoyancy extremum bound. Therefore~\eqref{PY112} tightly constrains $\la |\bomega|^2 \ra_{3D}$ to a narrow range.
For example, consider the implications of~\eqref{PY112} using the $\Ra = 6.4 \times 10^{13}$ \twoDFS{} solution in figures~\ref{Fig2}(c) and (d) as a comparison flow.
At $\Ra = 6.4 \times 10^{13}$  the 3D flow is not accessible to direct numerical solution.
But the 2D comparison solution has $\bar b_{2D}(0) / \bstar \approx -0.85$ and then~\eqref{PY112} becomes
\beq
  0.85 \frac{\kappa}{\nu} \leq \frac{\la |\bomega|^2\ra_{3D}}{\bstar/h}\leq \frac{\kappa}{\nu}\com \qquad \text{at $\Ra = 6.4 \times 10^{13}$.}
  \label{3D2D}
\eeq 
Thus at $\Ra = 6.4 \times 10^{13}$ the 3D flow has at most only 18\% more mean square vorticity than that of the 2D comparison flow in figure~\ref{Fig2}(d). 

With the argument above, numerical values of $\la |\bomega|^2 \ra$ and bottom buoyancy $b(0) / \bstar$ obtained from 2D direct numerical solution can be leveraged to constrain $\la |\bomega|^2 \ra$ of 3D HC flows at the same $\Ra$.
Moreover it is likely that as $\Ra \to \infty$ the 2D solutions will come closer to meeting the extremum bound. In this case as $\Ra \to \infty$, $\la |\bomega|^2 \ra_{3D}$ in~\eqref{PY112} is sandwiched into an ever narrower range and $\la |\bomega|^2 \ra \to \kappa \bstar/ \nu h$ for all values of $\elly / h$. 

In view of the controversial application of the zeroth law of turbulence to HC \citep{PY02, SW, Shish16} it is interesting that the efficacy of 3D HC vortex stretching  is   limited by the feeble supply of mechanical energy. 

As a concluding illustration of $\la |\bomega|^2 \ra \approx \kappa \bstar/ \nu h$, notice that in figures~\ref{Fig1} and~\ref{Fig2} the vorticity is scaled with $\sqrt{\bstar / h}$ even as $\Ra$ is varied by a factor of $10^6$.
Indeed this simple estimate of root mean square vorticity applies to all 2D and 3D solutions reported here, e.g., including the zoomed-in view of the BL in figure~\ref{Fig5}.
 
\section{Conclusion \label{conclusion2}}

We have conducted a numerical study of the $\Ra$--$\Nu$ relation with $\Pr = 1$ and four cases corresponding to no-slip or free-slip boundary conditions, in both 2D ($\elly / h = 0$) and 3D ($\elly / h = 1$) geometries.
In all four cases, with~$\Ra$ in the range $10^{6}$ to~$10^{10}$, we find that $\Nu \sim \Ra^{1/5}$.
In the 2D cases, with maximum Rayleigh number of order $10^{14}$, we find  a scaling regime with $\Nu \sim \Ra^{1/4}$; see also \cite{tsaiSheard20}. 

The $\Nu \sim Ra^{1/4}$ regime has a double boundary layer structure. There is a main boundary layer with thickness $\sim \Ra^{-1/5}$ and an inner boundary layer with thickness $\sim \Ra^{-1/4}$.
The thickness of the inner boundary layer corresponds to the Kolmogorov--Batchelor scales of HC in~\eqref{etaKB3}.

Scaling arguments for the $\Nu$--$\Ra$ relation of HC reviewed in section~\ref{RossbySec} do not depend very much, if at all, on the distinction between 2D and 3D HC.
Nor do these arguments identify the spanwise aspect ratio $\elly / h$ as an important parameter.
Thus it is informative to conduct parallel numerical studies of 2D and 3D HC and compare corresponding $\Nu$'s.
This comparison of 2D with 3D HC, extending to $\Ra = 3.2 \times 10^{11}$, shows that 3D HC has only a slight 10\% or 20\% enhancement of heat transport over non-turbulent 2D HC.
Boundary conditions are more important than dimensionality: the 2D free-slip solutions have larger $\Nu$ than 3D no-slip solutions at the same $\Ra$. 

The conclusions above are based on 3D computations with $\Ra \leq 3.2 \times 10^{11}$ and geophysical $\Ra$'s are larger by $10^{10}$ or $10^{12}$.
With the inequality~\eqref{PY112}, however, we can use results from 2D numerical solutions to constrain the mean square vorticity of 3D HC at the same $\Ra$.
With the additional plausible assumption that the bottom buoyancy of 2D solutions approaches that of the densest surface fluid, inequality~\eqref{PY112} implies that in 2D and in 3D the mean square vorticity of HC limits to $\kappa \bstar / \nu h$ as $\Ra \to \infty$. Thus vortex stretching and vorticity amplification do not operate in 3D HC.
These results reinforce the view that 3D HC does not express all of the characteristics of turbulence.

\section*{Acknowledgments}
Without implying their endorsement, we thank Basile Gallet, Ross Griffiths for discussions of horizontal convection.
Computational resources were provided by the Australian National Computational Infrastructure at the Australian National University, which is supported by the Commonwealth of Australia.
NCC was supported by the Australian Research Council DECRA Fellowship~DE210100749.
CBR was supported by National Aeronautics and Space Administration Award~NNX16AO5OH.
SGLS was partially supported by National Science Foundation Awards~OCE-1829919.
WRY was supported by National Science Foundation Awards~OCE-1657041 and OCE-2048583.

\section*{Declaration of interests}
The authors report no conflict of interest.

\appendix

\section{The low Rayleigh number regime \label{lowRa}}

If the Rayleigh number is sufficiently small then one can employ a straightforward expansion in powers of $\Ra$ to show that the Nusselt number is
\beq
  \Nu = 1 + C_2 \Ra^2 + \text{ord}\big( \Ra^4 \big)\per \label{lowRa1}
\eeq
In the expansion~\eqref{lowRa1}, $C_2$ is a function of the aspect ratio, $A_x = \ellx / h$, but not the Prandtl number $\Pr$.
In this appendix we summarize the calculation of $C_2$ for horizontal convection forced with the sinusoidal $\bs$ in~\eqref{sbuoy8}.
This calculation is more interesting than one might anticipate because $C_2$ turns out to be a very small number for all values of the aspect ratio~$A_x$.
Consequently with the aspect ratio $A_x = 4$ used in this work ``sufficiently small'' means Rayleigh numbers of order $10^4$ (see the inset in figure~\ref{Fig4}).

Using the streamfunction formulation, with $(u, w)=(-\psi_z, \psi_x)$, and scaling lengths with the depth $h$ and time with $h^2/\kappa$ the steady Boussinesq equations are
\begin{align}
  \psi_x \lap \psi_z - \psi_z \lap \psi_x & = b_x + \Pr \, \nabla^4 \psi \com \\
  \psi_x b_z - \psi_z b_x & = \lap b \com
\end{align}
where here $\lap = \p_x^2 + \p_z^2$ is the two-dimensional Laplacian; the spanwise vorticity is $u_z - w_x = - \lap \psi$.
The surface boundary condition is $b(x, 1) = \epsilon \cos mx$ where
\beq
  m \defn kh = {2\pi}/{A_x} \com \andd \epsilon \defn \Pr \Ra/A_x^3 = \Pr \Ra \, (m/2 \pi)^3 \per
  \label{epsilondef}
\eeq
We expand all variables in powers of the small parameter $\epsilon$
\beq
  (b, \psi) =\epsilon (b_1, \psi_1) + \epsilon^2 (b_2, \psi_2) + \cdots \per
\eeq

The first-order equations are
\begin{align}
  \Pr \nabla^4 \psi_1 = -b_{1x} \com \andd \lap b_1 = 0 \per
  \label{A66}
\end{align}
The solution~\eqref{A66} is
\beq
  \psi_1 = \sin m x \, P(z) \com \andd b_1 = \, \cos m x \, B(z) \com
\eeq
where $B(z) \defn \sech m \cosh m z$.
In $\psi_1(x, z)$ we have the free-slip function
\beq
  P^{\FS}(z) = \frac{B(z)}{8m^2 \Pr} \, \big[(m \coth m + 1 - z) \tanh mz - mz(2-z) \big]\com
\eeq
and the no-slip function
\begin{align}
  P^{\NS}(z) & = \frac{1}{8 m \Pr \, (\sinh^2m - m^2)} \bigg[ (\sinh^2m - m^2) (z^2-z) B(z) \nonumber \\
            & +(\tanh m - m)\, z \sinh m(1-z) + (\sinh m - m \sech m) \, (1-z) \sinh m z \bigg] \per
\end{align}

At second order in $\ep$ we must solve
\begin{align}
  \lap b_2 & = \psi_{1x} b_{1z} - \psi_{1z} b_{1x} \com \\
           & = \underbrace{\tfrac{1}{2} m (PB)'}_{J_0} \, + \, \underbrace{\tfrac{1}{2} m (PB' - P' B)}_{J_2}\, \cos 2 m x \per
  \label{lowRa11}
\end{align}
The solution of~\eqref{lowRa11} has the form
\beq
  b_2 = B_{20}(z) + B_{22}(z) \cos 2 m x\com
\eeq
where $B_{20}$ and $B_{22}$ are determined by solving
\begin{align}
  B''_{20} & = J_0\com \\
  B''_{22} - 4 m^2 B_{22}  & = J_2 \per
  \label{smallRa37}
\end{align}
Forming $\la B_{22} \eqref{smallRa37} \ra$ we find the shortcut used below in passing from~\eqref{smallRa47} to~\eqref{smallRa61}:
\beq
  \laa {B'_{22}}^2 + 4 m^2 B_{22}^2 \raa= - \laa B_{22} J_2\raa\per
\eeq
The expressions for $B_{22}$ and $B_{20}$, obtained with Mathematica, are complicated and are not explicitly presented.
Mercifully, to obtain the coefficient $C_2$ in~\eqref{lowRa1}, we do not need $\psi_2$.

Multiplying $ \lap b_1 = 0$ by $b_n$, with $n \geq 2$, and noting that all these $b_n$'s have homogeneous boundary conditions at $z = 0$ and $1$, we see that $\la \grad b_n \bcdot \grad b_1 \ra =0$.
Consequently the expansion of the buoyancy variance dissipation is
\begin{align}
  \chi &= \epsilon^2 \!\!\! \underbrace{\la |\grad b_1|^2 \ra}_{\chi_2 = \half m \tanh m} \!\!\!+ \; \epsilon^4 \underbrace{\la |\grad b_2|^2 \ra}_{\chi_4} \, + \, \text{ord} \big(\epsilon^6 \big) \per
  \label{smallRa47}
\end{align}
Recalling the definition of $\epsilon$ in~\eqref{epsilondef}, the Nusselt number is
\beq
  \Nu = 1 + \underbrace{\frac{m^2 \big\la (PB)^2\big\ra - 2 \la B_{22} J_2\ra}{2m \tanh m}}_{\chi_4 / \chi_2 }\, \left(\frac{m^3 \Pr}{8 \pi^3} \right)^2 \Ra^2 + \cdots \per
  \label{smallRa61}
\eeq
Because $P$, $B_{22}$, and $J_2$ are all proportional to $\Pr^{-1}$, the Prandtl number $\Pr$ cancels out of the coefficient of $\Ra^2$ in~\eqref{smallRa61}.

With \FS{} boundary conditions, the expression for $C_2$ in~\eqref{lowRa1} is
\begin{align}
  C^{\FS}_2= & \Big[690 - 1920 m^4 \cosech^2m + 20 (3 + 4m^2)^2 \sech 2m \nonumber \\
  & \quad + m(1024m^4 - 80 m^2 -6195)\, \cosech m \, \sech^3 m \nonumber \\
  & \quad \quad +5(352 m^4 - 624 m^2 + 1065) \sech^2m \Big] \Big \slash 41\, 943 \, 040 \, \pi^6 \per
  \label{lowRa23}
\end{align}
Limiting values in the \FS{} case are
\beq
  \lim_{m \to 0} C^{\FS}_2 = \frac{31 \, m^8}{30 \, 965 \, 760 \pi^6} + \text{ord} \big(m^{10} \big) \com
  \label{smnostress}
\eeq
and
\beq
  \lim_{m \to \infty} C^{\FS}_2 = \frac{69}{4\, 194\, 304\, \pi^6}\per
\eeq
We admire the frequently occurring integer $4\, 194 \, 304=2^{22}$ in the formulas above and below.

With \NS{} boundary conditions, we find
\beq
  C^{\NS}_2= \frac{1}{41\, 943\, 040 \, \pi^6} \frac{\Xi(m) }{\big( \cosh 2 m -2 m^2 - 1 \big)^2} \com
\eeq
where
\begin{align}
  \Xi(m) & = 85 \cosh 4m - 320 m(33 + 8 m^2) \sinh 2 m + 10\big( 845 + 716 m^2\big) \cosh 2 m \nonumber \\
  & \ - 5 (5123 + 18304 m^2 + 13720 m^4 + 2912 m^6) \nonumber \\
  & \ \ - 4m^5(7215 + 100m^2 - 64 m^4)\coth m \nonumber \\
  & \ \ \ + 20 (829 + 4402 m^2 + 5289 m^4 + 836 m^6 - 64 m^8) \sech^2m \nonumber \\
  & \ \ \ \ - 4 m (4785 + 8010 m^2 + 1549 m^4 - 212 m^6 + 64 m^8) \tanh m \sech^2 m\nonumber \\
  & \ \ \ \ \ + 20 (25 - 184 m^2 - 80 m^4 - 16 m^6) \sech 2 m \nonumber \\
  & \ \ \ \ \ \ + 4 m (8865 + 21640 m^2 + 17719 m^4 + 740 m^6 - 64 m^8) \tanh m\nonumber \\
  & \ \ \ \ \ \ \ +160 m (15 - 7 m^2 - 4 m^4) \tanh 2 m \per
  \label{lowRa24}
\end{align}
Limiting values in the \NS{} case are
\beq
  \lim_{m\to 0}C^{\NS}_2 = \frac{m^8}{30\, 965\, 760 \pi^6} + \text{ord}\big(m^{10}\big)\com
  \label{smnoslip}
\eeq
and
\beq
  \lim_{m \to \infty} C^{\NS}_2 = \frac{17}{4\, 194\, 304 \pi^6} \per
\eeq
It is notable that the small-$m$ $C^{\NS}_2$ in~\eqref{smnoslip} is smaller by a factor of exactly 31 than $C^{\FS}_2$ in~\eqref{smnostress}.

Both $C_2$'s are very much less than one for all aspect ratios.
In the numerical solutions summarized in figure~\ref{Fig4}, the aspect ratio is $A_x = 4$, corresponding to $m = k h = \pi / 2$.
The results in~\eqref{nstud3} and~\eqref{nstud4} follow by evaluating the formulas~\eqref{lowRa23} through~\eqref{lowRa24} with $m = \pi/2$.


\begin{thebibliography}{31}
  \expandafter\ifx\csname natexlab\endcsname\relax\def\natexlab#1{#1}\fi
  \def\au#1{#1} \def\ed#1{#1} \def\yr#1{#1}\def\at#1{#1}\def\jt#1{\textit{#1}} \def\bt#1{#1}\def\bvol#1{\textbf{#1}} \def\vol#1{#1} \def\pg#1{#1} \def\publ#1{#1}\def\arxiv#1{#1}\def\org#1{#1}\def\st#1{\textit{#1}}
  
  \bibitem[Benzi {\em et~al.\/}(1987)Benzi, Patarnello \& Santangelo]{Benzi1987}
  {\sc \au{Benzi, R}, \au{Patarnello, S} \& \au{Santangelo, P}} \yr{1987}  \at{On the statistical properties of two-dimensional decaying turbulence}.  \jt{Europhys. Lett.}  \bvol{3}~(7),  \pg{811}.
  
  \bibitem[Benzi {\em et~al.\/}(1988)Benzi, Patarnello \& Santangelo]{Benzi1988}
  {\sc \au{Benzi, R}, \au{Patarnello, S} \& \au{Santangelo, P}} \yr{1988}  \at{Self-similar coherent structures in two-dimensional decaying turbulence}.  \jt{J.~Phys. A}  \bvol{21}~(5),  \pg{1221}.
  
  \bibitem[Burns {\em et~al.\/}(2020)Burns, Vasil, Oishi, Lecoanet \& Brown]{Dedalus2020}
  {\sc \au{Burns, K.~J.}, \au{Vasil, G.~M.}, \au{Oishi, J.~S.}, \au{Lecoanet, D.} \& \au{Brown, B.~P.}} \yr{2020}  \at{Dedalus: {A} flexible framework for numerical simulations with spectral methods}.  \jt{Phys. Rev. Res.}  \bvol{2},  \pg{023068}.
  
  \bibitem[Carnevale {\em et~al.\/}(1991)Carnevale, McWilliams, Pomeau, Weiss \& Young]{Carnevale1991}
  {\sc \au{Carnevale, G.~F.}, \au{McWilliams, J.~C.}, \au{Pomeau, Y.}, \au{Weiss, J.~B.} \& \au{Young, W.~R.}} \yr{1991}  \at{Evolution of vortex statistics in two-dimensional turbulence}.  \jt{Phys. Rev. Lett.}  \bvol{66}~(21),  \pg{2735}.
  
  \bibitem[Chiu-Webster {\em et~al.\/}(2008)Chiu-Webster, Hinch \& Lister]{CWHL08}
  {\sc \au{Chiu-Webster, S.}, \au{Hinch, E.~J.} \& \au{Lister, J.~R.}} \yr{2008}  \at{Very viscous horizontal convection}.  \jt{J.~Fluid Mech.}  \bvol{611},  \pg{395--426}.
  
  \bibitem[Coman {\em et~al.\/}(2006)Coman, Griffiths \& Hughes]{CGH06}
  {\sc \au{Coman, M.~A.}, \au{Griffiths, R.~W.} \& \au{Hughes, G.~O.}} \yr{2006}  \at{Sandstr\"om's experiments revisited}.  \jt{J.~Mar. Res.}  \bvol{64}~(6),  \pg{783--796(14)}.
  
  \bibitem[Dritschel {\em et~al.\/}(2008)Dritschel, Scott, Macaskill, Gottwald \& Tran]{Dritschel2008}
  {\sc \au{Dritschel, D.~G.}, \au{Scott, R.~K.}, \au{Macaskill, C.}, \au{Gottwald, G.~A.} \& \au{Tran, C.~V.}} \yr{2008}  \at{Unifying scaling theory for vortex dynamics in two-dimensional turbulence}.  \jt{Phys. Rev. Lett.}  \bvol{101}~(9),  \pg{094501}.
  
  \bibitem[Gayen {\em et~al.\/}(2014)Gayen, Griffiths \& Hughes]{G14}
  {\sc \au{Gayen, B.}, \au{Griffiths, R.~W.} \& \au{Hughes, G.~O.}} \yr{2014}  \at{Stability transitions and turbulence in horizontal convection}.  \jt{J.~Fluid Mech.}  \bvol{751},  \pg{698--724}.
  
  \bibitem[Hughes \& Griffiths(2008)]{HG08}
  {\sc \au{Hughes, G.~O.} \& \au{Griffiths, R.~W.}} \yr{2008}  \at{Horizontal convection}.  \jt{Annu. Rev. Fluid Mech.}  \bvol{40},  \pg{185--208}.
  
  \bibitem[Ilicak \& Vallis(2012)]{IV12}
  {\sc \au{Ilicak, M.} \& \au{Vallis, G.~K.}} \yr{2012}  \at{Simulations and scaling of horizontal convection}.  \jt{Tellus~A}  \bvol{64}~(1),  \pg{18377}.
  
  \bibitem[McWilliams(1984)]{McW1984}
  {\sc \au{McWilliams, J.~C}} \yr{1984}  \at{The emergence of isolated, coherent vortices in turbulent flow.}  \jt{J.~Fluid Mech.}  \bvol{146},  \pg{21--43}.
  
  \bibitem[McWilliams(1990)]{McW1990}
  {\sc \au{McWilliams, J.~C}} \yr{1990}  \at{The vortices of two-dimensional turbulence}.  \jt{J.~Fluid Mech.}  \bvol{219},  \pg{361--385}.
  
  \bibitem[Mullarney {\em et~al.\/}(2004)Mullarney, Griffiths \& Hughes]{M04}
  {\sc \au{Mullarney, J.~C.}, \au{Griffiths, R.~W.} \& \au{Hughes, G.~O.}} \yr{2004}  \at{Convection driven by differential heating at a horizontal boundary}.  \jt{J.~Fluid Mech.}  \bvol{516},  \pg{181--209}.
  
  \bibitem[Paparella \& Young(2002)]{PY02}
  {\sc \au{Paparella, F.} \& \au{Young, W.~R.}} \yr{2002}  \at{Horizontal convection is non-turbulent}.  \jt{J.~Fluid Mech.}  \bvol{466},  \pg{205--214}.
  
  \bibitem[Passaggia {\em et~al.\/}(2017)Passaggia, Scotti \& White]{PSW17}
  {\sc \au{Passaggia, Pierre-Yves}, \au{Scotti, Alberto} \& \au{White, Brian}} \yr{2017}  \at{Transition and turbulence in horizontal convection: linear stability analysis}.  \jt{J. Fluid Mech.}  \bvol{821},  \pg{31--58}.
  
  \bibitem[Ramme \& Hansen(2019)]{RH2019}
  {\sc \au{Ramme, Lennart} \& \au{Hansen, Ulrich}} \yr{2019}  \at{Transition to time-dependent flow in highly viscous horizontal convection}.  \jt{Physical Review Fluids}  \bvol{4}~(9),  \pg{093501}.
  
  \bibitem[Reiter \& Shishkina(2020)]{ReiterShish2020}
  {\sc \au{Reiter, Philipp} \& \au{Shishkina, Olga}} \yr{2020}  \at{Classical and symmetrical horizontal convection: detaching plumes and oscillations}.  \jt{J.~Fluid Mech.}  \bvol{892}.
  
  \bibitem[Rocha {\em et~al.\/}(2020{\natexlab{{\em a\/}}})Rocha, Bossy, Llewellyn~Smith \& Young]{RBLSY}
  {\sc \au{Rocha, C.~B.}, \au{Bossy, T.}, \au{Llewellyn~Smith, S.~G.} \& \au{Young, W.~R.}} \yr{2020{\natexlab{{\em a\/}}}}  \at{Improved bounds on horizontal convection}.  \jt{J. Fluid Mech.}  \bvol{883},  \pg{A41}.
  
  \bibitem[Rocha {\em et~al.\/}(2020{\natexlab{{\em b\/}}})Rocha, Constantinou, Llewellyn~Smith \& Young]{RocNussDef}
  {\sc \au{Rocha, C.~B.}, \au{Constantinou, N.~C.}, \au{Llewellyn~Smith, S.~G.} \& \au{Young, W.~R.}} \yr{2020{\natexlab{{\em b\/}}}}  \at{The {N}usselt numbers of horizontal convection}.  \jt{J.~Fluid Mech.}  \bvol{894},  \pg{A24}.
  
  \bibitem[Rossby(1965)]{R65}
  {\sc \au{Rossby, H.~T.}} \yr{1965}  \at{On thermal convection driven by non-uniform heating from below: an experimental study}.  \jt{Deep Sea Res.}  \bvol{12}~(1),  \pg{9--10, IN9--IN14, 11--16}.
  
  \bibitem[Rossby(1998)]{R98}
  {\sc \au{Rossby, H.~T.}} \yr{1998}  \at{Numerical experiments with a fluid heated non-uniformly from below}.  \jt{Tellus~A}  \bvol{50}~(2),  \pg{242--257}.
  
  \bibitem[Sandstr\"om(1908)]{S08}
  {\sc \au{Sandstr\"om, J.~W.}} \yr{1908}  \at{Dynamische versuche mit meerwasser}.  \jt{Annals in Hydrodynamic Marine Meteorology}  \bvol{36},  \pg{6--23}.
  
  \bibitem[Scotti \& White(2011)]{SW}
  {\sc \au{Scotti, A.} \& \au{White, B.}} \yr{2011}  \at{Is horizontal convection really ``non-turbulent?''}.  \jt{Geophys. Res. Lett.}  \bvol{38}~(21),  \pg{L21609}.
  
  \bibitem[Sheard \& King(2011)]{SK2011}
  {\sc \au{Sheard, G.~J.} \& \au{King, M.~P.}} \yr{2011}  \at{Horizontal convection: effect of aspect ratio on {Rayleigh} number scaling and stability}.  \jt{Appl. Math. Modelling}  \bvol{35}~(4),  \pg{1647--1655}.
  
  \bibitem[Shishkina {\em et~al.\/}(2016)Shishkina, Grossmann \& Lohse]{Shish16}
  {\sc \au{Shishkina, O.}, \au{Grossmann, S.} \& \au{Lohse, D.}} \yr{2016}  \at{Heat and momentum transport scalings in horizontal convection}.  \jt{Geophys. Res. Lett.}  \bvol{43}~(3),  \pg{1219--1225}.
  
  \bibitem[Shishkina \& Wagner(2016)]{SW16}
  {\sc \au{Shishkina, Olga} \& \au{Wagner, Sebastian}} \yr{2016}  \at{Prandtl-number dependence of heat transport in laminar horizontal convection}.  \jt{Physical Review Letters}  \bvol{116}~(2),  \pg{024302}.
  
  \bibitem[Siggers {\em et~al.\/}(2004)Siggers, Kerswell \& Balmforth]{SKB04}
  {\sc \au{Siggers, J.~H.}, \au{Kerswell, R.~R.} \& \au{Balmforth, N.~J.}} \yr{2004}  \at{Bounds on horizontal convection}.  \jt{J.~Fluid Mech.}  \bvol{517},  \pg{55--70}.
  
  \bibitem[Tsai {\em et~al.\/}(2016)Tsai, Hussam, Fouras \& Sheard]{Tsai16}
  {\sc \au{Tsai, T.}, \au{Hussam, W.~K}, \au{Fouras, A.} \& \au{Sheard, G.~J.}} \yr{2016}  \at{The origin of instability in enclosed horizontally driven convection}.  \jt{Int. J.~Heat and Mass Transfer}  \bvol{94},  \pg{509--515}.
  
  \bibitem[Tsai {\em et~al.\/}(2020)Tsai, Hussam, King \& Sheard]{tsaiSheard20}
  {\sc \au{Tsai, T.}, \au{Hussam, W.~K.}, \au{King, M.~P.} \& \au{Sheard, G.~J.}} \yr{2020}  \at{Transitions and scaling in horizontal convection driven by different temperature profiles}.  \jt{Int. J.~Thermal Sci.}  \bvol{148},  \pg{106166}.
  
  \bibitem[Wang \& Huang(2005)]{WH2005}
  {\sc \au{Wang, W.} \& \au{Huang, R.~X.}} \yr{2005}  \at{An experimental study on thermal convection driven by horizontal differential heating}.  \jt{J.~Fluid Mech.}  \bvol{540},  \pg{49--73}.
  
  \bibitem[Winters \& Young(2009)]{WY09}
  {\sc \au{Winters, K.~B.} \& \au{Young, W.~R.}} \yr{2009}  \at{Available potential energy and buoyancy variance in horizontal convection}.  \jt{J.~Fluid Mech.}  \bvol{629},  \pg{221--230}.
  
  \end{thebibliography}
\end{document}